\documentclass[%
 reprint,
superscriptaddress,
nofootinbib,
 amsmath,amssymb,mathtools
 aps,
]{revtex4-2}

\usepackage{color,soul}
\usepackage{subfigure}

\usepackage{breqn}
\usepackage[dvipsnames]{xcolor}
\usepackage{graphicx}
\usepackage{dcolumn}
\usepackage{bm}
\usepackage{hyperref}

\usepackage[ruled,vlined,linesnumbered]{algorithm2e}
\SetKwInput{KwInput}{Input} 
\DontPrintSemicolon
\usepackage{cases}
\usepackage{enumerate}					

\makeatletter
\let\cat@comma@active\@empty
\makeatother

\begin{document}

\preprint{APS/123-QED}

\title{Decoding probabilistic syndrome measurement and the role of entropy}

\author{Jo\~{a}o F. Doriguello}
\affiliation{Centre for Quantum Technologies, National University of Singapore, Singapore}
\affiliation{School of Mathematics, University of Bristol, Bristol, United Kingdom}
\affiliation{Quantum Engineering Centre for Doctoral Training, University of Bristol, Bristol, United Kingdom}

\date{\today}

\begin{abstract}

In realistic stabiliser-based quantum error correction there are many ways in which real physical systems deviate from simple toy models of error. Stabiliser measurements may not always be deterministic or may suffer from erasure errors, such that they do not supply syndrome outcomes required for error correction. In this paper, we study the performance of the toric code under a model of probabilistic stabiliser measurement. We find that, even under a completely continuous model of syndrome extraction, the threshold can be maintained at reasonably high values of $1.69\%$ by suitably modifying the decoder using the edge-contraction method of Stace and Barrett (Physical Review A 81, 022317 (2010)), compared to a value of $2.93\%$ for deterministic stabiliser measurements. Finally, we study the role of entropic factors which account for degenerate error configurations for improving on the performance of the decoder. We find that in the limit of completely continuous stabiliser measurement any advantage further provided by these factors becomes negligible in contrast to the case of deterministic measurements.

\end{abstract}

\maketitle


\section{\label{sec:intro}Introduction}

To achieve scalable quantum computation, quantum error correction is required to address unavoidable noise on physical qubits. Quantum error-correcting codes~\cite{kitaev2003fault,dennis2002topological} can encode quantum information and, combined with a decoder, can enable fault-tolerant computation despite the existence of errors on physical qubits. There are useful benchmarking methods to analyse the performance of error-correcting codes, such as using simple toy error models which abstract away many of the details of a physical system that would actually be used to implement such a code. However, to make progress towards quantum error correction in practice, it is important to analyse the performance of codes and decoders when features of a hardware system are reintroduced. Here we isolate and analyse one such feature which deviates from a simple toy error model: asynchronous measurement. 

To perform quantum error correction, parity check measurements are repeatedly made on the qubits of the code. A usual setting in topological codes is that these measurement are made deterministically in `rounds', i.e., on demand (not necessarily error-free), such that in each round every qubit of the code is involved in one parity check. However, when any of the operations that are used to perform a parity check are non-deterministic, this assumption does not apply. Parity checks may be inherently probabilistic, as is the case when they depend on ancillary states from non-deterministic entanglement generation or distillation procedures, e.g.\ in modular quantum-computing architectures~\cite{Monroe,nickerson2015practical,kim2011modular}. In other systems, parity checks may be subject to measurement erasure, where measurement outcomes are not always returned, e.g.\ in photonic quantum computing~\cite{knill2001scheme} where single-photon detectors suffer optical loss~\cite{Mercedes,Morley_Short_2019,Morley_Short_2017}. 

In this work we study a model of asynchronous parity check measurement in the toric code. In this model the stabiliser measurements are attempted at discrete times and each attempt provides a parity outcome with probability $s$, called the \emph{synchronicity} parameter. We push this to the limit $s\to 0$ where parity checks are performed continuously. For an independent and identical distributed (i.i.d.)\ error model and a minimum-weight perfect matching (MWPM) decoder~\cite{edmonds1967optimum,kolmogorov2009blossom,Davies1998}, the toric code exhibits a threshold of $2.93\%$ when parity checks are entirely synchronous~\cite{wang2003confinement}. We show that, by marking unsuccessful parity checks as erased in the syndrome graph (the `history' of stabiliser measurement outcomes), it is possible to contract erased edges in the syndrome graph into multi-edges following the method described by Stace and Barrett~\cite{stace2010error}. This gives a clear framework on how to properly include non-identical error probabilities arising from asynchronism into a MWPM decoder which, when appropriately modified, can maintain the threshold at a reasonably high value of $1.69\%$ in the completely continuous regime.

A secondary motivation for studying this model is to explore the role of degeneracy in the MWPM decoder under asynchronous measurements. It is known~\cite{duclos2010fast,stace2010error} that accounting for degeneracy, i.e., the number of shortest paths that are consistent with a matching, can improve the usual MWPM decoder's threshold: for an i.i.d.\ error model with faultless, fully synchronous ($s=1$) stabiliser measurements, path counting boosts the MWPM decoder threshold from $10.3\%$ to $10.65\%$~\cite{stace2010error}. Moreover, degeneracy has also been used to close the gap between minimum-weight perfect matching and optimal methods~\cite{criger2018multi}, as well as to compare different variants of the toric code with a comparable number of qubits~\cite{beverland2019role}. Here we study how to introduce degeneracy into the MWPM decoder under asynchronism by considering multi-path counting on top of the edge-contraction method, and we observed a mild improvement from $1.69\%$ to $1.70\%$ on the decoder's threshold. We argue, and provide numerical evidence, that the presence of asynchronism increases the predominance, i.e., the relative probability, of the most likely error configuration over all the others, thus diminishing the role of degeneracy on decoding.

Sec.~\ref{sec:sec2} reviews the toric code and introduces our toy model of asynchronism. Sec.~\ref{sec:sec4} discusses the approach to decoding and the way in which degeneracy appears. It also introduces our proposed decoding algorithms. Their performance is then benchmarked in Sec.~\ref{sec:sec9}. We further discuss our results and conclude in Sec.~\ref{sec:sec7}.


\begin{figure*}[t]
    \centering
    \includegraphics[width=0.9\textwidth]{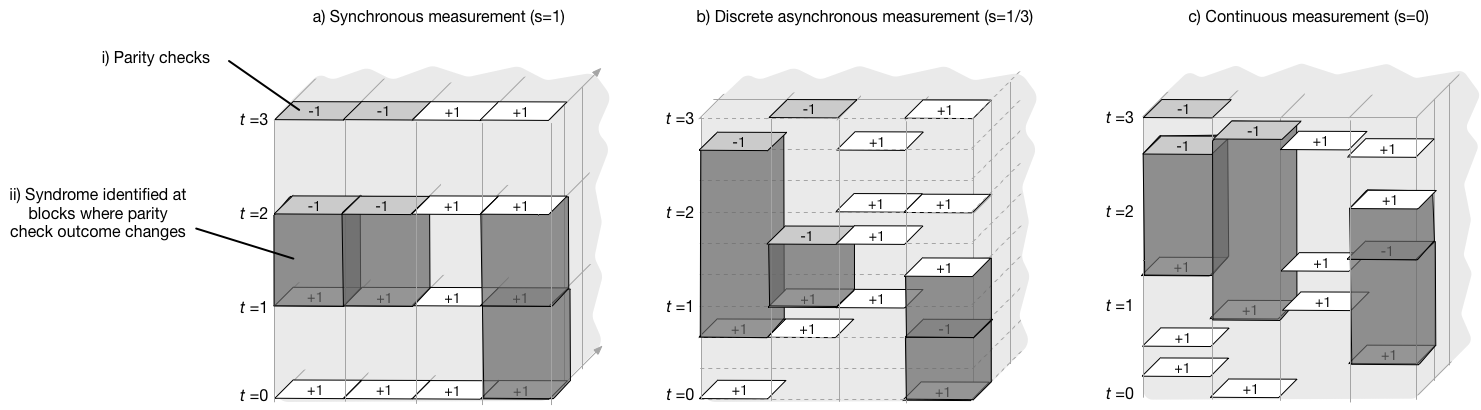}
    \caption{Illustration of three regimes of asynchronous stabiliser measurement in the square toric code. (a) Fully synchronous ($s=1$). Repeated measurement on the code created a 3-dimensional block of parity check outcomes over time. Outcomes are measured deterministically in layers at discrete time intervals. An anyon (a violated syndrome bit) is identified when a parity check measurement changes from one round to the next (dark grey blocks). A physical error will result in two anyons separated in space. A measurement error will result in two anyons separated in time. (b) Discrete asynchronous measurement ($0 < s < 1$). Measurements are performed in discrete rounds, but an outcome is only returned with probability $s$, where in the figure $s=\frac{1}{3}$. Physical or measurement errors result in a pair of anyons, but these are now identified in intervals of varying size as indicated in the figure. (c) Continuous asynchronous ($s=0$): parity check measurements can happen at any time, at a rate 1 per unit time.}
    \label{fig:asynchronous_illustration}
\end{figure*}

\section{Asynchronism in the toric code}
\label{sec:sec2}

\subsection{The toric code}

The toric code~\cite{kitaev2003fault} is a topological code defined on an $L\times L$ square lattice with periodic boundary condition, where a qubit is located on each edge of the lattice. There is an operator $X_v$ and $Z_f$ associated with each vertex $v$ and each face $f$ of the lattice, respectively. The code space is defined as the simultaneous `$+1$' eigenstate of the operators $X_v$ and $Z_f$. $X_v$ is the product of the Pauli-$X$ matrices acting on edges incident to $v$, i.e., $X_v = \prod_{e \ni v} X_e$, while $Z_f = \prod_{e \in f} Z_e$ is the product of the Pauli-$Z$s acting on all edges of the face $f$. These operators, and any product of them, form the stabiliser group, $S$. Logical operators are made up of $X$ and $Z$ operators acting on a string of qubits that span the lattice, giving rise to logical operators $\overline{X}_1$ and $\overline{Z}_2$ along one direction, and $\overline{X}_2$ and $\overline{Z}_1$ along the other direction. 
To achieve fault tolerance the stabiliser operators $X_v$ are measured. If there is an error $E_Z \in \{I, Z\}^{\otimes n}$, any stabiliser $X_v$ that anticommutes with the error returns a `$-1$' outcome. To account for the fact that the stabiliser measurements themselves can also be subject to error, the stabilisers are measured multiple times, and parity check operators are defined as the product of two subsequent measurements of the same stabiliser generator. If no error occurs during both measurements, then the parity check will return a `$+1$' outcome. If a Pauli error occurs between the first and second measurement, or if there is a measurement error in one of the measurements, then the parity check will return a `$-1$' outcome, which can be seen as a quasi-particle and is called \emph{anyon}. The subset of parity checks with `$-1$' measurement outcomes is called the {\em syndrome} $\sigma$. Given a syndrome $\sigma$, a decoder can then be applied to find a correction operator $\mathcal{C}(\sigma)$ such that $\mathcal{C}(\sigma) E_Z \in S$. That is, if the correction operator is applied to the code, the error is corrected up to a stabiliser. We note that in quantum computation it is not necessary to physically apply any correction operator to the qubits, rather the correction can be thought of as a reference frame through which the measurement outcomes can be interpreted.

\subsection{Asynchronous stabiliser measurement}
\label{sec:asyn_stab_meas}

We now introduce a model of asynchronous stabiliser measurement. This model is designed to isolate the effects of measurement asynchronicity while leaving all other features of the system the same. But it is worth highlighting that there could be many things about this model that could be changed depending on the physical system. 

Consider a square toric code of size $L \times L$. The toric code is subject to repeated measurements for a time $T$. Each attempted stabiliser measurement provides a parity outcome with probability $s$, a parameter we called the {\em synchronicity} of the system. Otherwise, with probability $1-s$, no outcome is obtained, which is marked as a `$0$' outcome, i.e., erased. Parity measurements are successfully recorded at a rate $1$ per unit time on average, meaning that $1/s$ measurements are attempted in one unit of time. We define two measures of errors on qubits: the \emph{simulation error} $p_\Delta$ and the \emph{physical error} $p$. The simulation error is the probability that a qubit suffers an error between two consecutive parity check attempts. The physical error is the probability that a qubit suffers an error per unit time, i.e., after $1/s$ parity check attempts. The physical and simulation errors are related as follows: the probability that a qubit suffers an error after $n$ measurement rounds equals the probability that during these $n$ rounds its state is flipped an odd number of times (each with probability $p_\Delta$), i.e.,
\begin{align}
    \sum_{m~\text{odd}}^{n} \binom{n}{m}p_\Delta^m (1-p_\Delta)^{n - m} = \frac{1}{2}\left(1 - (1 - 2p_\Delta)^n\right).\label{eq:eq1.1}
\end{align}
Since a time unit represents $1/s$ measurement rounds on average, both quantities $p$ and $p_\Delta$ are related via
\begin{subequations}
\label{eq:eq2.1}
\begin{align}
    p &= \frac{1}{2}\big(1 - (1 - 2p_\Delta)^{1/s}\big), \label{eq:eq2.1a}\\
    p_\Delta &= \frac{1}{2}\big(1 - (1 - 2p)^{s}\big).\label{eq:eq2.1b}
\end{align}
\end{subequations}
Finally, successful measurements are subject to measurement errors, which flip the outcome value with probability $q=p$.

When $s=1$, measurements are fully synchronous. When $s\to0$, measurements are completely continuous. By fixing the rate $p$ of physical errors and the rate of successful parity checks (set to $1$), we are able to probe the behavior of the code with respect to the parameter $s$. We consider three distinct regimes, which are illustrated in Fig.~\ref{fig:asynchronous_illustration}: 

\begin{enumerate}
    \item {\bf Synchronous measurement ($s=1$).} This corresponds to the error model with fully synchronous parity checks.
    \item {\bf Discrete asynchronous measurement ($0<s<1$).} Measurements are performed in discrete rounds, but are not deterministic and occur with probability $s$. Measurement rounds are performed at a rate $1/s$ such that the overall rate of successful stabiliser measurement remains at $1$ per unit time. 
    \item {\bf Continuous measurement ($s=0$).} Measurements are not performed in rounds, but are received continuously at a rate $1$. Similarly, Pauli errors are treated as continuous. The times of the successful measurements and Pauli errors are modelled as arising from a Poisson distribution, the resulting distribution obtained from the Binomial distribution in the limit $s\to0$ (see Appendix~\ref{app:app_sim}).
\end{enumerate}

One can see from Fig.~\ref{fig:asynchronous_illustration} the effect of the probabilistic nature of parity checks. Successful stabiliser measurements are separated in time, thus creating a block-like structure. Every stabiliser operator $X_v$ has an ordered list of measurement times for successful parity checks $(t^v_1,t^v_2,\dots)$, where $t^v_1 < t^v_2 < \dots$. Two consecutive measurement times define a \emph{parity block}. More specifically, the $i$th parity block associated with $v$ is defined by the pair of time coordinates $(t^v_{i-1},t^v_i)$. If the measurement outcomes differ from each other at consecutive times $t^v_{i-1}$ and $t^v_{i}$, then we refer to this block as an \emph{anyon block}. In the fully synchronous regime ($s=1$), two consecutive measurements with differing outcomes lead to an anyon well defined in time. On the other hand, for $s<1$, such anyons (now anyon blocks) are spread over time. Defining their time position is one of the main issues in constructing the decoding problem and correcting for errors.

\begin{figure*}
    \centering
    \includegraphics[width=2\columnwidth]{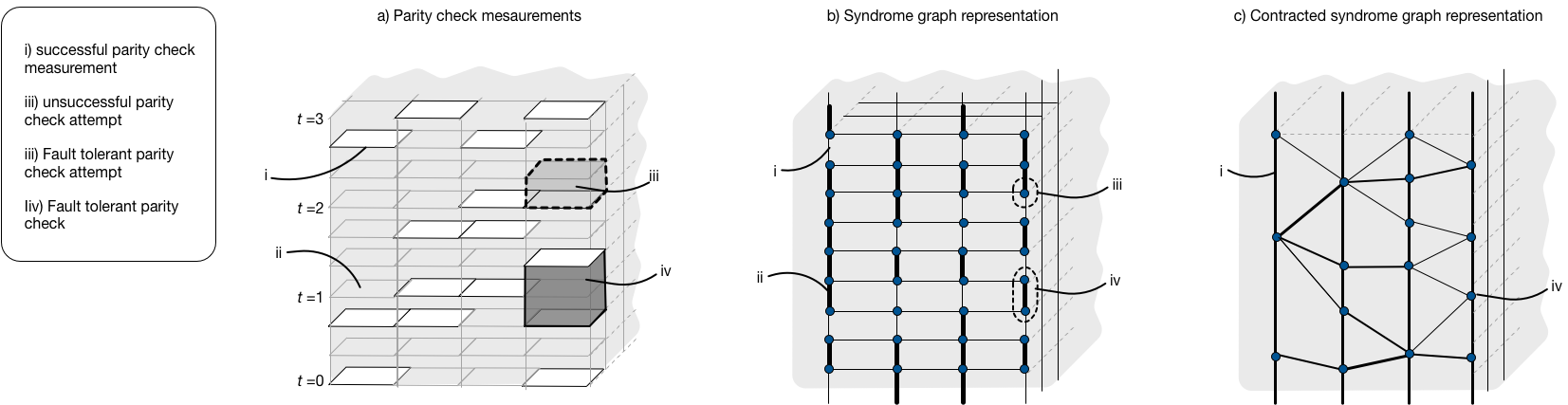}
    \caption{Constructing the contracted syndrome graph in the case of discrete asynchronous stabiliser measurement. We follow the method described first in~\cite{stace2010error}. (a) Collection of all parity check attempts through time, including successful and unsuccessful measurements. Two consecutive successful parity checks in time define a parity block. (b) Simple syndrome graph representation. Horizontal edges represent a possible physical Pauli error and vertical edges represent an attempted stabiliser measurement outcome. A vertical edge with unsuccessful parity check is marked as erased (bold edge). (c) The contracted syndrome graph representation. All vertices and vertical edges within a parity block are contracted into a single vertex. Horizontal edges connecting adjacent parity blocks are contracted into a single edge and new edge weights are calculated in order to reflect the degeneracy of the new contracted edges. \label{fig:decoding_graph}
    }
\end{figure*}

\subsection{Constructing the decoding problem}

To analyse fault tolerance in this system we first want to formulate the error model and structure of the code as a {\em syndrome graph}.
In the syndrome graph, vertices represent fault-tolerant parity checks and edges represent the potential errors in the system, e.g.\ physical and measurement errors. This representation is the most useful way to analyse the performance of decoding algorithms as it fully describes the system, capturing both space and time behavior. 

Each edge in the syndrome graph is assigned a bit that indicates whether or not an error has occurred. 
Vertices are assigned a parity value which is computed as the parity of the values of all edges incident to that vertex. If there are no errors, all vertices will have an even parity. If an error occurs, the two vertices connected to the corresponding edge will have their values flipped.

For a fault-tolerance system there may be multiple possible syndrome graph representations that capture the same error model. We consider first the {\em simple syndrome graph} that is most naturally derived from the parity check structure. We then study the {\em contracted syndrome graph}. 

\subsubsection{Simple syndrome graph} 

When all parity measurements are performed synchronously, the syndrome graph has a cubic structure. Time-like edges represent the possible measurement errors on parity checks with error probability $q=p$, while space-like errors represent potential Pauli errors on the physical qubits with error probability $p_\Delta=p$. As previously mentioned, the set of all odd parity checks defines the syndrome and two consecutive parity checks with differing outcomes define an anyon in between both measurement times.

In our model of asynchronous measurement, we have to modify this representation since not all parity checks return an outcome. This is done by marking an edge of the graph as erased when there is a corresponding measurement erasure. The net result is that multiple sequential erasures in time lead to a `block' of marked edges. The formulation of this system into a syndrome graph, named simple syndrome graph, is illustrated in Fig.~\ref{fig:decoding_graph}. Space-like edges are still associated with error probability $p_\Delta$ and non-erased time-like edges with error probability $q=p$. As previously mentioned, anyons are no longer well defined in time, as the `blocks' of erased edges can now have variable time lengths. We note that the graph structure, i.e., its cubic structure, is the same in the every instance, the only difference being the position of the erased edges.


\subsubsection{Contracted syndrome graph}

Given a simple syndrome graph with a set of erased edges as shown in Fig.~\ref{fig:decoding_graph}(b), we find an alternative representation without erased edges. When erasure is present, fault-tolerant parity checks are only complete for each cluster of erased edges~\cite{stace2010error}. In our case this means simply treating all the vertices between two successful measurements as one vertex, i.e., considering a parity block as a vertex. By contracting the graph around the erased edges, we arrive at the contracted syndrome graph. An example is shown in Fig.~\ref{fig:decoding_graph}(c). The contraction resolves the problem of defining the anyons. These are now placed at contracted vertices that are between two consecutive parity checks with differing outcomes, or, in other words, at contracted vertices associated with anyon blocks.

Carrying out the contraction will often result in multi-edges in the graph, where two erased components were connected by multiple edges in the simple syndrome graph. These correspond to multiple possible errors that could cause the same syndrome. An equivalent representation that is more convenient for decoding is to instead represent these as a single edge with modified error probability. Such modified error probability is related to the physical error $p$ and the time overlap of erased components in the simple syndrome graph, as illustrated in Fig.~\ref{fig:decoding_graph}, via the same reasoning that relates $p_\Delta$ and $p$ in Eq.~\eqref{eq:eq2.1}. More specifically, let $\omega_{ij} =  \min(t_i,t_j) - \max(t_{i-1},t_{j-1})$ be the time overlap between two adjacent parity blocks $(t_{i-1},t_i)$ and $(t_{j-1},t_j)$. The number of merged edges is therefore just $\omega_{ij}/s$. The probability $\bar{p}(\omega_{ij})$ of a Pauli error occurring on the merged edge of the contracted syndrome graph is equal to the probability of a Pauli error occurring an odd number of times on the corresponding edges from the simple syndrome graph, which is given by Eq.~(\ref{eq:eq1.1}):
\begin{align}
    \label{eq:eq4A.1b}
    \bar{p}(\omega_{ij}) = \frac{1 - (1 - 2p_\Delta)^{\omega_{ij}/s}}{2} = \frac{1 - (1 - 2p)^{\omega_{ij}}}{2}.
\end{align}
%
A merged edge between two adjacent parity blocks with time overlap $\omega(e)$ has thus an associated error probability $\bar{p}(\omega(e))$. Time-like (vertical) edges continue to represent possible measurement errors with probability $q = p$. The resulting contracted syndrome graph thus offers a simple and compact framework in which decoding techniques can be straightforwardly used. We will show how to apply such decoding techniques in the following section.


\subsubsection{Continuous stabiliser measurement}

In the case of continuous stabiliser measurement when $s=0$, there is no way to construct the simple syndrome graph. In this case we build the contracted syndrome graph directly by recording the parity check measurement times. Given the locations of successful parity checks, a vertex is identified with each parity block. An edge is then placed between vertices whose adjacent parity blocks overlap in time and has an associated probability according to Eq.~(\ref{eq:eq4A.1b}).

\section{Decoding}
\label{sec:sec4}

The job of the decoder is to identify a {\em correction} for a given syndrome graph, in the form of a predicted set of flipped edges. An optimal decoder identifies corrections that minimise the chance of logical errors. For practical use however, efficient decoding algorithms are required that approximate  optimal decoding while being computationally tractable~\cite{hutter2014efficient,duclos2010fast,delfosse2017almost}. In this section, we describe the decoding strategies for probabilistic syndrome measurement that will be analysed in Sec.~\ref{sec:sec9}.

\subsection{Anyon-pairing decoders}
A correction in the toric code can be expressed as a pairing of anyons (odd-parity check vertices). Any two error chains that produce the same syndrome but differ by trivial cycles have the same effect on the logical state. The task that should be performed by a decoder can be understood as matching anyons in a way that minimises the chance of a logical error. A large range of decoders can be defined as performing  minimum-weight perfect matching (MWPM)~\cite{cook1999computing} on \emph{matching graphs} derived from syndrome graphs. A matching graph is specified for any syndrome graph as a \emph{complete} graph where the vertices correspond to the anyons, and the edge weights between vertices correspond to distances in the syndrome graph. It is then necessary to properly weight the edges in the syndrome graph, which we now describe.

When constructing a matching graph, we would ideally like to compute each \emph{anyon pairing probability}, i.e., the probability that \emph{any} error created an observed anyon pair. Performing decoding on a matching graph with these probabilities then reveals the \emph{most likely pairing} of anyons, and is independent of the type of syndrome graph (e.g.\ simple or contracted). In other words, we would like to compute the pairing probability between anyons $i$ and $j$ as given by
\begin{align}
    P_{ij} = \sum_{E \in \mathbb{E}} P_E  = P_0 + P_1 + P_2 + \dots,  \label{eq:Pij_iid}
\end{align}
where $E$ is an error chain (a set of odd parity outcomes) whose boundaries are the vertices $i$ and $j$, $P_E$ is its probability, and $\mathbb{E}$ is the set of all such error chains. The error chains are indexed $E=0,1,2,\dots$ from most to least likely.

Consider now an error model where each edge $e$ in the syndrome graph represents an independent (but not necessarily identical) error occurring with probability $p_e$. The probability for each error chain can be expressed as
\begin{align*}
    P_E = \prod_{e \in E }p_e \prod_{e\notin E } (1-p_e) &=  C \prod_{e \in E}\frac{p_e}{1-p_e},
\end{align*}
where $C = \prod_{\forall e} (1-p_e)$ is a constant for a given syndrome graph.  Commonly, a simplified MWPM decoder is used that identifies only most likely errors for the correction operator, corresponding to approximating $P_{ij}$ by $P_0$ for each anyon pairing in Eq.~(\ref{eq:Pij_iid}).  Since
\begin{align}
    \label{eq:p0}
    \ln{P_0} = \operatorname*{\max}_E \ln{P_E} = \ln{C} - \operatorname*{\min}_E \sum_{e \in E}\ln\left(\frac{1-p_e}{p_e}\right),
\end{align}
the error chains that are used must minimise $\sum_{e \in E}\ln((1-p_e)/p_e)$. The distance between any two pairs of anyons to be inputted into the matching graph can be found by using Dijkstra's algorithm~\cite{dijkstra1959note} on the syndrome graph with edges weighted by $\ln((1-p_e)/p_e)$. The minimisation itself, i.e., the anyon pairing with overall minimum additive weight, can be found via Edmond's minimum-weight, perfect-matching algorithm~\cite{edmonds1967optimum}.



A MWPM decoder can be improved by considering more terms $P_E$ in the anyon pairing probability. In the usual fully synchronous ($s=1$) i.i.d.\ error model with $q=p$ (equal physical and measurement error probabilities), $P_E = Cp^{| E|}$, where $|E|$ is the length of the error chain. It is then possible that two or more paths have the same probability $P_E$, meaning they are degenerate. The question is thus reduced to counting the number of paths with a given length between two anyons. The introduction of degeneracy for the shortest path into the MWPM decoder, i.e., considering all terms $P_E$ equal to $P_0$ in the pairing probability, was examined in~\cite{stace2010error}. When erasure is present and we work with the contracted syndrome graph, it is important to introduce a suitable notion of degeneracy for error chains when estimating anyon pairing probabilities. Recall that the probability $p_e$ of an edge $e$ in the contracted syndrome graph is given by $p_e = \frac{1}{2}(1 - (1 - 2p)^{\omega(e)})$, where $\omega(e)$ is the time overlap between the blocks defining $e$ (see Eq.~\eqref{eq:eq4A.1b}). By approximating $p_e \approx p\cdot \omega(e)$ and assuming $p_e$ is small, we find
%
%
%
\begin{align}
    P_{ij} &= C \sum_{E \in \mathbb{E}} \prod_{e \in E}\frac{p_e}{1-p_e}\nonumber\\
    &\approx C \sum_{E \in \mathbb{E}} \prod_{e \in E}\frac{p\omega(e)}{1-p\omega(e)}\nonumber\\
    &\approx C \sum_{E \in \mathbb{E}} \prod_{e \in E}(p\omega(e) + (p\omega(e))^2)\nonumber\\
    &= C \sum_{E \in \mathbb{E}} \left(\prod_{e \in E}p\omega(e)\right)\left(\prod_{e\in E}(1 + p\omega(e))\right)\nonumber\\
    &\approx C \sum_{E \in \mathbb{E}} p^{|E|} \delta_E,\label{eq:6.3approximation2}
\end{align}
where in Eq.~\eqref{eq:6.3approximation2} we defined the quantity $\delta_E = \prod_{e\in E}\omega(e)$ (which is the product of the $\omega(e)$ values along the error chain), and we approximated $\prod_{e\in E}(1 + p\omega(e))\approx 1$. This last approximation can be justified as follows. Time overlaps $\omega(e)$ are expected to be smaller than $1$ on average (since parity blocks have length $1$ on average), so we can write $\prod_{e\in E}(1 + p\omega(e)) \lessapprox e^{p|E|}$.  For large chains, if the lattice size $L$ is sufficiently large so that $L$ is comparable to $p^{-1}$, $e^{p|E|} \approx e^{pL}$ might be considerable, and the actual probability of large chains is underestimated. However, for small chains, the approximation $\prod_{e\in E}(1 + p\omega(e))\approx 1$ is fairly accurate, and these are the ones that are relevant to the decoder since the most likely error configurations are typically composed of small error chains. The probability underestimation for large chains is thus ignored by the decoder.

By grouping terms for which error chains have the same number of edges, we obtain the following expression for the pairing probability:  
\begin{align}
    \label{eq:eq_f1}
    P_{ij} \propto \sum_{l\geq l_0} p^{l}\sum_{E\in\mathbb{E}_l}\delta_E,
\end{align} 
where $l_0$ denotes the length of the shortest error chain connecting $i$ and $j$, and $\mathbb{E}_{l}$ is the set of error chains connecting $i$ and $j$ of length $l$.
We define the $(k+1)$th-order {\em degeneracy factor} for each term in $p^{l}=p^{l_0+k}$ to be
\begin{align}
    \Omega_{k} := \sum_{E\in\mathbb{E}_{l_0+k}} \delta_E.\label{eq:eq_f2}
\end{align}
These factors are closely related to counting the number of paths with the same {\em number} of errors, i.e., edges. Indeed, for the i.i.d.\ error model with synchronicity $s=1$, $\delta_E = 1$ for all paths, and so $\Omega_k$ equals exactly the number of paths with $l_0+k$ edges (see more in Appendix~\ref{app:appA0}). 

\subsection{Decoding algorithms}


In this section we propose several decoders based on different approximations for $P_{ij}$.

\subsubsection{Contracted Syndrome Graph decoders}

We first consider a MWPM decoder on the contracted syndrome graph with the approximation $P_{ij} = P_0$, which we name Contracted Graph (CG) decoder. 
The probability $P_0$ is given by Eq.~\eqref{eq:p0} with $p_e = \bar{p}(\omega(e)) = \frac{1}{2}(1-(1-2p)^{\omega(e)})$. Finding the most likely error chain is equivalent to $\min_E \sum_{e\in E}\ln\left((1-p_e)/p_e\right)$ and, hence, the CG decoder weights each edge by $\ln((1-p_e)/p_e)$ and proceeds to find the path with the minimum additive weight. In other words, this weight assignment defines a metric $d_C$ in the contracted syndrome graph. Therefore, the weight between two anyon blocks $i$ and $j$ is set as the shortest distance between them,
\begin{align}
    w_{ij} = d_C(i,j).
    \label{eq:eq6B.2}
\end{align}

The CG decoder can be enhanced by keeping more terms in Eq.~\eqref{eq:Pij_iid}. From Eq.~\eqref{eq:eq_f1} we can keep the first two groups of terms with shortest lengths ($\mathbb{E}_{l_0}$ and $\mathbb{E}_{l_0+1}$) with their corresponding degeneracy terms  $\Omega_{0}$ and $\Omega_{1}$. Similarly to Eq.~\eqref{eq:p0}, the CG decoder should now optimise $\max_E\ln{P_{ij}} \propto \max_E\ln\big(p^{|E|}\Omega_0 + p^{|E|+1}\Omega_1\big) = -\min_E\big[|E| \ln{p^{-1}} - \ln(\Omega_0 + p\cdot \Omega_1)\big]$. Therefore, the weight assignment between a pair of anyons $i$ and $j$ is
\begin{align}
    \label{eq:eq6.C1}
    w_{ij} &= l_0(i,j)\ln{p^{-1}} - \tau\ln\left(\Omega_{0} + p\cdot\Omega_{1}\right)
\end{align}
up to an additive constant, and where we included a parameter $\tau$, named \emph{degeneracy parameter}, that can be tuned in order to improve the decoder performance. 
Efficient computation of degeneracies $\Omega_{0}$ and $\Omega_{1}$ can be done via Dijkstra's algorithm, as explained in Appendix~\ref{app:appA}.


\subsubsection{Approximated decoders}


One of the drawbacks of the CG decoder is the lack of a closed-form expression for the distance between two anyons in the metric $d_C$, since erased time-like edges are randomly distributed. This means that we must use Dijkstra's algorithm to compute such distances, which can be too slow for the situation at hand. It is thus interesting to propose heuristic approximations to the CG decoder that do not require the use of Dijkstra's algorithm and have a close-form expression for the distance between two anyon blocks given their coordinates. In order to do so, we work with the simple syndrome graph given its cubic structure.

\begin{figure}[t]
    \centering
    \includegraphics[trim={3.5cm 18cm 5cm 4.5cm},clip,width=0.8\textwidth]{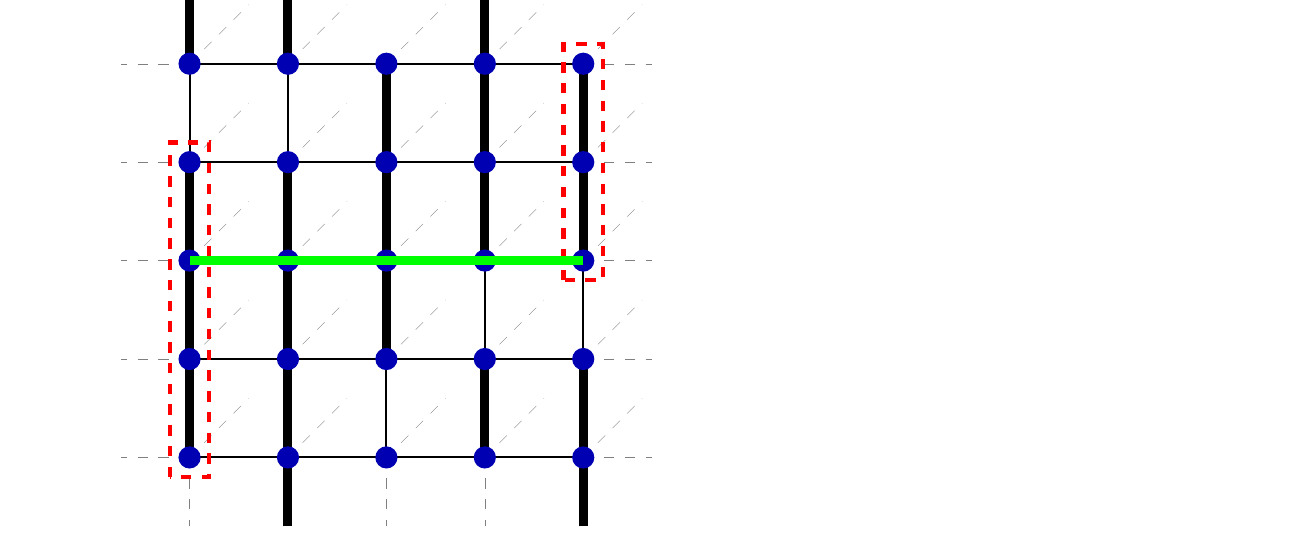}
    \caption{Two anyon blocks, highlighted in red, that overlap in time can be matched by a minimum-length error chain with no time-like edges, shown in green. The time distance between both blocks is zero.}
    \label{fig:overlap-blocks}
\end{figure}

For our first approximation, we treat the anyon blocks as defined anyons in a fully synchronous simple syndrome graph, thus ignoring erased edges. This means taking a Manhattan distance between two anyon blocks as their weight into the matching graph. More specifically, consider two anyon blocks with coordinates $(x_i,y_j,t_{i1},t_{i2})$ and $(x_j,y_j,t_{j1},t_{j2})$. Their spacial distance is $\Delta(x_i,x_j) + \Delta(y_i,y_j)$, where $\Delta (x_i,x_j) = \min(x_i-x_j~(\text{mod}~L), x_j-x_i~(\text{mod}~L))$ is the $x$ horizontal distance on the lattice (and similarly for the $y$ coordinate). Moreover, if both blocks overlap in time, then their time distance is zero, since there is an error chain with minimal length with no time-like edges connecting both blocks (see Fig.~\ref{fig:overlap-blocks}). If the blocks do not overlap in time, we take the average number of non-erased time edges between them. Suppose, e.g.\ that $t_{i1} > t_{j2}$. There are $(t_{i1}-t_{j2})/s$ time-like edges between both blocks, out of which $t_{i1}-t_{j2}$ are non-erased on average. We thus can approximate their time distance as $\max(t_{i1}-t_{j2}, t_{j1}-t_{i2}, 0)$ (note this is $0$ when the blocks overlap in time, since $t_{i1}-t_{j2}$ and $t_{j1}-t_{i2}$ are negative). Given these considerations, we propose the Block Graph (BG) decoder which, instead of finding the actual distances within the contracted syndrome graph using Dijkstra's algorithm, sets the distance (and thus the weight) between two anyon blocks with coordinates $(x_i,y_j,t_{i1},t_{i2})$ and $(x_j,y_j,t_{j1},t_{j2})$ as
\begin{multline}
    \label{eq:app-simple}
    w_{ij} = \Delta(x_i,x_j) + \Delta(y_i,y_j) \\
    + w_{\rm time}^{\rm BG}\max(t_{i1}-t_{j2},t_{j1}-t_{i2},0),
\end{multline}
where we introduced a tunable parameter $w^{\rm BG}_{\rm time}$. The weight between anyons in the matching graph becomes then a function of only their coordinates.

Our second approximation within the simple syndrome graph is to reduce the analysis back to a cubic syndrome graph by defining a specific time coordinate for each anyon. More specifically, each anyon is identified at a time location in the middle of its corresponding anyon block. For example, if an anyon block is defined by times $t_{i-1}$ and $t_{i}$, then the corresponding anyon is given a location $(t_{i}+t_{i-1})/2$. Our proposed Average Position (AP) decoder treats these anyons as existing in a cubic syndrome graph, and computes the weight between two anyons $i$ and $j$ with coordinates $(x_i,y_i,t_i)$ and $(x_j,y_j,t_j)$ using the Manhattan distance,
\begin{equation}
    w_{ij} = \Delta (x_i,x_j) + \Delta (y_i,y_j) + w^{\rm AP}_{\rm time} |t_i - t_j|,
    \label{eq:eq1}
\end{equation}
where we introduced a tunable parameter $w^{\rm AP}_{\rm time}$.

\section{Results}
\label{sec:sec9}

\subsection{Simulation methods}
\label{sec:sec_sim}

To study the performance of each decoding algorithm we perform Monte Carlo simulations of the system where errors are sampled and the resulting system is decoded and analysed to determine whether or not an error is introduced. Since we consider a model of independent $X$ and $Z$ errors we directly simulate only phase-flip errors and $X$-type parity checks, as by symmetry the performance will be the same for bit-flip errors. For each decoding algorithm we simulate its performance for a range of stabiliser synchronicity $s\in[0,1]$. Our simulations capture both discrete probabilistic measurements, where stabiliser measurements are made at discrete time steps with varying success probability, and continuous measurement for which we sample errors and measurements over a continuous range. Here we briefly outline the simulation methods for both for these cases, and more details on simulation techniques can be found in Appendix~\ref{app:app_sim}.

\subsubsection{Threshold performance}

\textbf{Discrete Measurement.} To model discrete probabilistic stabiliser measurements we sample measurements and errors on the simple cubic syndrome graph. We define a time scale such that stabiliser measurements are obtained at a rate $1$ after $1/s$ time steps on average. In other words, each measurement round is performed after a time interval $s$. At each time step each physical qubit (space-like edge) suffers a flip with probability $p_\Delta$, the simulation error, where $p_\Delta$ is related to the physical error rate $p$ (error probability after $1/s$ time steps) via $p_\Delta = \frac{1}{2}\big(1 - (1 - 2p)^{s}\big)$ (Eq.~\eqref{eq:eq2.1}). Each stabiliser measurement (time-like edge) is sampled and is successfully measured with probability $s$. If the measurement does not succeed then the edge is marked as erased, otherwise, if it does succeed, then its value is flipped with probability $q$, the measurement error. We take $q=p$.

\textbf{Continuous measurement.} To model continuous stabiliser measurement ($s=0$) we cannot directly sample stabiliser measurements as probabilistic events. Instead we sample error events and measurement events over a continuous time period, aiming to keep all the error parameters equivalent to the discrete measurement case. We set a time interval $T$ and a physical error $p$ per unit time. For each qubit we sample the number of bit-flips it suffers in the time interval $T$ from a Poisson distribution with parameter $\frac{T}{2} \ln(1/(1-2p))$ (see Appendix~\ref{app:app_sim} for a justification). Given the number of events we then sample their time coordinate from a uniform distribution along the time interval $(0,T)$. This gives us a set of space-time coordinates of qubit flip events. For each stabiliser site we sample the number of successful measurements from a Poisson distribution with parameter $T$ and distribute these measurements uniformly at random along the time interval $(0,T)$. This gives us a set of space-time coordinates for stabiliser measurements. A parity check is done by counting the number of errors of the adjacent qubits prior to the measurement time. If the number is even (odd), the measurement outcome is $+1$ ($-1$). For faulty measurements this outcome is flipped with probability $q = p$.
 
Given the locations of parity check measurements we then directly construct the contracted syndrome graph by identifying a vertex with each successive pair of parity checks, and edges between neighboring check locations where parity blocks have a non-zero time overlap. Each edge has an associated error probability $p_e = \bar{p}(\omega(e))$, where $\omega(e)$ is the time overlap of the parity blocks defining the edge $e$ (Eq.~\eqref{eq:eq4A.1b}).

\textbf{Parameter optimisation.} The AP decoder, the BG decoder and the CG decoder augmented with $\Omega_0$ and $\Omega_1$ have tunables parameters, to know, the time and degeneracy parameters $w^{\rm AP}_{\rm time}$, $w^{\rm BG}_{\rm time}$ and $\tau$, respectively. For a given value of $s$, we probe their dependence on these parameters and pick the optimum value when comparing the threshold performance between different decoders. Their dependence on $w^{\rm AP}_{\rm time}$, $w^{\rm BG}_{\rm time}$ and $\tau$ is explored in Appendix~\ref{app:appA0}.

\subsubsection{Analysing entropic contributions}

In addition to gauging the decoders' performance via their threshold, we want to understand how good an approximation is being made to the anyon pairing probability. To do this we compute the average magnitude $\langle P_E/P_0\rangle$ of the first few terms $P_E$ from Eq.~\eqref{eq:Pij_iid} relative to the zeroeth-order term $P_0$. If the higher-order terms have small values, then we expect the proposed decoders to perform well. To obtain these ratios we perform a further series of numerical experiments. We fix the synchronicity $s$, a physical error $p$ and a measurement error $q=p$, and, by sampling errors via Monte Carlo simulation as previously described, we obtain a syndrome and identify all pairs of anyons that \emph{are matched by the decoder} (not any pair of anyon). The ratio $P_E/P_0$ is then computed for each such \emph{matched} pair using Yen's algorithm~\cite{yen1971finding}, which is a generalisation of Dijkstra's algorithm for computing the $k$-shortest loopless paths in a graph with non-negative edge cost. We average this value over all the matched anyon pairs and, finally, over other random contracted syndrome graphs.

\subsection{Threshold performance} 

\begin{figure*}[t]
 \begin{center}
    \subfigure[Threshold dependence with synchronicity $s$ for Contracted Graph (CG), Block Graph (BG) and Average Position (AP) decoders.]{\label{fig:fig3}\includegraphics[width=0.49\textwidth]{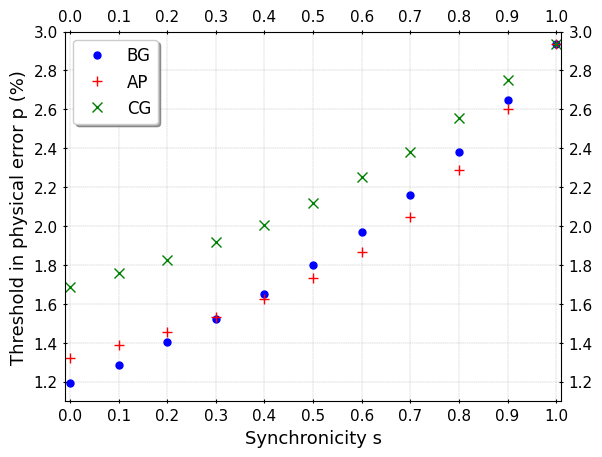}}
    \subfigure[Threshold dependence with synchronicity $s$ for Contracted Graph (CG) decoder with and without the degeneracy terms $\Omega_0$ and $\Omega_1$.]{\label{fig:fig4}\includegraphics[width=0.49\textwidth]{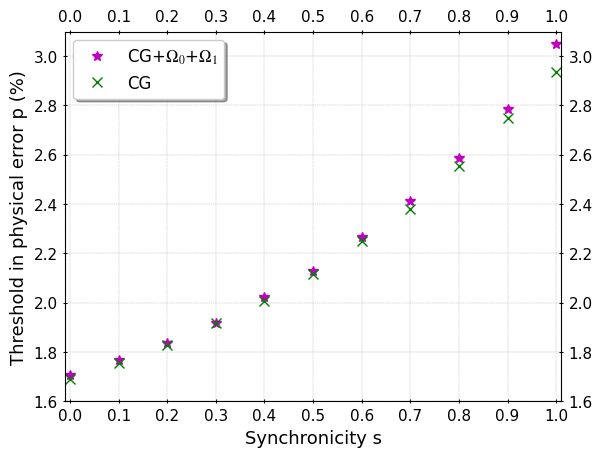}}
    \end{center}
    \caption{Threshold comparison between all decoders. The points were calculated using an $L\times L$ lattice with $L\in\{10,12,14\}$ and $N_s$ measurement rounds with $N_s = \left\lceil 2/s\right\rfloor L$ for $s\in(0,1]$, and $N_0=T=2L$ for $s=0$.}\label{fig:fig4a}
\end{figure*}

Fig.~\ref{fig:fig4a} shows our main results, the threshold performance with synchronicity $s\in[0,1]$ for the three main different decoders introduced in the previous section. 
Fig.~\ref{fig:fig3} compares all decoders. On the other hand, Fig.~\ref{fig:fig4} specifically compares the CG decoder with and without the degeneracy terms $\Omega_0$ and $\Omega_1$. At $s=1$ we have the usual MPMW decoder for the toric code with faulty measurements and i.i.d.\ error model~\cite{wang2003confinement}, hence all decoders perform identically. As the synchronicity $s$ decreases, the performance of all decoders decreases, as expected. Nonetheless, even at the limit of continuous stabiliser measurement ($s=0$), the threshold can be maintained at a reasonably high level, e.g.\ $1.688\%\pm 0.001\%$ for the CG decoder. On the other hand, the simplification of the syndrome graph structure by the BG and AP decoders, while speeding up the decoding procedure, leads to a decrease in threshold values, e.g.\ $1.20\%\pm 0.01\%$ (BG decoder) and $1.32\%\pm 0.01\%$ (AP decoder) at $s=0$. Interestingly, the AP decoder, even though inferior to the BG decoder for high values of $s$, outperforms it for high asynchronism, a fact for which we do not have an explanation. In Appendix~\ref{app:appA0} we show more information on the AP and BG decoders, e.g.\ their dependence on the time parameters $w_{\rm time}^{\rm AP}$ and $w_{\rm time}^{\rm BG}$ at $s=0$ and the optimal values for $w_{\rm time}^{\rm AP}$ and $w_{\rm time}^{\rm BG}$ as a function of asynchronicity. In addition, we also show how the inclusion of degeneracy terms like $\Omega_0$ and $\Omega_1$ into the BG decoder can lead to a substantial threshold increase.

Something that stands out from Fig.~\ref{fig:fig4} is the fact that, while the introduction of high-order degeneracy like $\Omega_{1}$ does give higher threshold values compared to the base case of the CG decoder, this improvement becomes very small in the limit $s\to 0$. Even by $s=0.9$ the reduction is significant. While at $s=1$ the threshold increases from $2.937\%$ to $3.050\%$ (an $\sim 0.11\%$ additive improvement), at $s=0$ it only increases from $1.688\%$ to $1.699\%$ (an $\sim 0.01\%$ additive improvement). This feature is not entirely surprising, given the following. If one assumes that the set of possible error probabilities $\{p_e\}_e$ on each edge is very diverse, e.g.\ consider the case of continuous asynchronism where $p_e = \bar{p}(\omega(e))$ and $\omega(e)$ can be any real number in $(0,T)$, then it becomes very unlikely to have two degenerate error chains. Therefore, for completely different error probabilities $\{p_e\}_e$, we expect most of the terms in Eq.~\eqref{eq:Pij_iid} to be different from each other. This is in contrast to the fully synchronous regime ($s=1$), where most first terms are equal (see more in Appendix~\ref{app:appA0}). Consequently, the leading term $P_0$ plays a more prominent role in the sum, and any truncation to it is less disruptive to its original value when $s=0$ compared to when $s=1$. 

\begin{figure}
    \centering
    \includegraphics[width=\columnwidth]{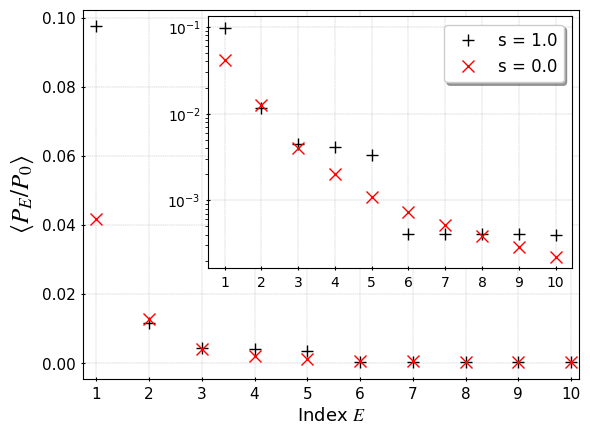}
    \caption{Average ratio $\langle P_E/P_0\rangle$ as a function of $E$ in the fully synchronous ($s=1$) and continuous asynchronous ($s=0$) regimes for an $L\times L\times 2L$ lattice with $L=14$. The inner figure is in logarithmic scale. The ratios were averaged over matched anyon pairs by the CG decoder and over random contracted syndrome graphs given a physical error $p$ and a measurement error $q=p$. We chose $p=2.937\%$ and $p=1.688\%$ for $s=1$ and $s=0$, respectively, which are the thresholds of the CG decoder from Fig.~\ref{fig:fig4a}.}
    \label{fig:fig9}
\end{figure}

In order to support the above claim, we shed some light on the relative size between the first $P_E$ terms and $P_0$ which underlies the decrease in threshold values. Fig.~\ref{fig:fig9} explores how much smaller the first few terms in $\sum P_E$ are in comparison with $P_0$ both in the fully synchronous ($s=1$) and continuous asynchronous ($s=0$) regimes. The average ratio $\langle P_E/P_0\rangle$ is obtained for $E=1,\dots,10$. One can see that most of high-order contribution is coming from $P_1$ on average, which is also where the discrepancy between high and low synchronicity regimes lies: $P_1$'s contribution is more than double in the regime $s=1$ compared to its contribution when $s=0$. On the other hand, asynchronism has a much smaller impact on the average $\langle P_E/P_0\rangle$ for high-order terms $E>1$. The relative importance of $P_1$ over other high-order terms $P_E$ is explicitly shown in Appendix~\ref{app:appA0} for the fully synchronous i.i.d.\ error model.

\subsection{Advantage in logical gate time}
    
\begin{figure}
    \centering
    \includegraphics[width=\columnwidth]{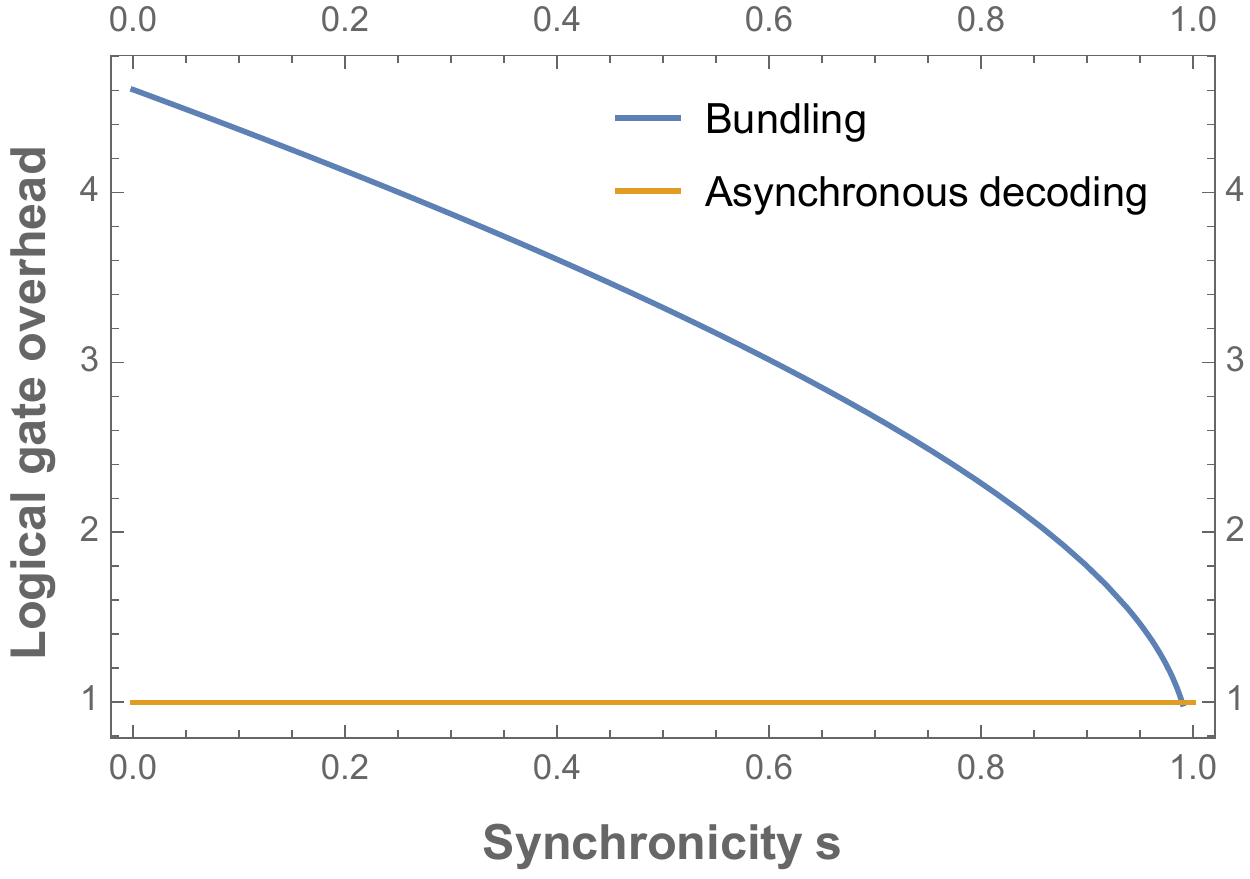}
    \caption{Comparing logical gate execution times of the bundling method described in~\cite{nickerson2013topological} with a renormalised synchronicity $s'=0.99$ to our asynchronous decoding approach.}
    \label{fig:bundle_vs_async}
\end{figure}

Direct handling of asynchronous stabiliser measurement in decoding can also provide an advantage in the time needed to execute logical gates. In Ref.~\cite{nickerson2013topological}, probabilistic stabiliser measurements arise in a scheme for quantum computing using networked ion traps. In this situation the physical errors occur only during successful stabiliser measurement, and so there is no penalty to the threshold for lower measurement probability. Ref.~\cite{nickerson2013topological} handles the probabilistic nature of the measurements by waiting for as many attempts as necessary to get to $99\%$ success across all stabiliser sites, and abandoning the remaining $1\%$, whose impact on the threshold is negligible. This essentially redefines a `round' of stabiliser measurement to be made up of $N_R$ rounds, such that $1-s' = (1-s)^{N_R}<0.01$. Once a stabiliser site is successfully measured it idles and waits until the round is completed. The cost to this approach is in the time taken to execute logical operations. In the limit of small success probability many attempts must be taken to complete a renormalised round, and during this time many of the sites spend a significant time idling. We can define the logical gate overhead, $R_L$, as the ratio between the number of rounds needed to complete renormalised round vs the number needed to measure a stabiliser on average as
\begin{align*}
    R_L = \frac{N_R}{1/s} =  \frac{\log{(1-s')}}{\log{(1-s)}} s
\end{align*}
for $s<0.99$, and $R_L = 1$ for $s\geq 0.99$. By using asynchronous decoding, each stabiliser is measured independently of the others, allowing stabiliser measurement information to be gathered at a faster rate, and giving $R_L=1$. Fig.~\ref{fig:bundle_vs_async} shows that, for high asynchronism, the bundling method from~\cite{nickerson2013topological} takes more than four times on average to perform a round compared to our asynchronous decoding approach. Moreover, their bundling method, which presupposes that physical errors occur due to successful stabiliser measurements, would probably not be viable in a more stringent error model like ours where physical errors can happen in between stabiliser measurements.

\section{Conclusions}
\label{sec:sec7}

We have shown how asynchronism can be incorporated into MWPM decoders while still maintaining a high threshold. We considered a simple error model where a stabiliser measurement outputs an outcome with probability $s$ called synchronicity. The limit $s\to 0$ represents a continuous asynchronous regime were stabilisers are measured completely at random in time. We tackled asynchronism by marking unsuccessful stabiliser measurements as erased in the simple syndrome graph, followed by contracting each cluster of erased edges using an edge-contraction method from Stace and Barrett~\cite{stace2010error}. The resulting graph was named contracted syndrome graph and, in opposition to the simple one, offers an easy framework to take non-identical error probabilities into consideration when decoding. We then proposed a MWPM-like decoder, named Contracted Graph (CG), using a properly weighted contracted syndrome graph.

We benchmarked the CG decoder via Monte Carlo and observed that the threshold values do decrease as the synchronicity tends to zero, but a significant level can be maintained even under a completely continuous model of syndrome extraction, e.g.\ the CG decoder holds thresholds of $1.69\%$ at $s=0$. While our results were obtained with a simple error model, they show that erasure errors suffered by measurements can be efficiently handled by decoders. Studying the performance of the CG decoder under more realistic error models is a point to be considered in the future.

The CG decoder is relatively simple: by being a MPWM decoder, one only requires performing Dijkstra's algorithm on a rightly weighted graph. However, even though being a polynomial-time algorithm, it could be too slow for practical applications. Indeed, running Dijkstra's algorithm requires time $O(|E| + |V|\log{|E|})$, where $|E|$ and $|V|$ are the number of edges and vertices of the syndrome graph, respectively. The syndrome graph is fairly sparse ($|E| = O(|V|)$), meaning that each application of Dijkstra's algorithm takes $\widetilde{O}(|V|) = \widetilde{O}(L^3)$ time in the contracted syndrome graph (the notation $\widetilde{O}$ hides polylog factors). Since Dijkstra's algorithm must be used once for each of the $O(pL^3)$ anyons, this leads to the overall time complexity $\widetilde{O}(pL^6)$.

In order to remedy this, we proposed the Block Graph (BG) and Average Position (AP) decoders that skip any use of Dijkstra's algorithm by approximating the distance between two anyons, which can be calculated in constant time. The overall time complexity improves to $O(p^2L^6)$. However, the price is a decrease in threshold value down to $1.32\%$ at $s=0$, which is still reasonably high. Given the simple structure of these decoders, specially the AP, it might be possible to borrow previous techniques used to improve the basic MWPM decoder~\cite{kolmogorov2009blossom,fowler2012topological} (some of these ideas could possibly be applied to the CG decoder as well). Finally, the AP and BG decoders allow for the introduction of auxiliary parameters like the time weights $w_{\rm time}^{\rm AP}$ and $w_{\rm time}^{\rm BG}$, which must be tweaked depending on the error model. Understanding their performance as a function of $w_{\rm time}^{\rm AP}$ and $w_{\rm time}^{\rm BG}$ with more mathematical rigour is something that we did not tackle and should be considered in the future.

Another aspect we explored was the role of degeneracy terms under asynchronous measurements and how they could be included into the decoder. More specifically, we study the inclusion of the first and second-order degeneracy terms $\Omega_{0}$ and $\Omega_{1}$ into the CG decoder. Such inclusion only produced a mild improvement in threshold, from $1.69\%$ to $1.70\%$ in the limit $s\to 0$, which hints to the fact that the role of degeneracy becomes less important in the continuous asynchronous regime and considering only the lowest weight error configuration becomes an increasingly better approximation. This was further backed up by our numerical results on the relative size between the most likely error configurations. We showed that, as the synchronicity decreases, the probability of the most likely error configuration becomes relatively higher than the probability of the subsequent ones. In might be interesting to understand this behaviour in a more qualitative manner, although it might be a hard task given the similarity to the problem of counting trails.


\section*{Acknowledgments}

We would like to specially thank Hugo Cable and Naomi Nickerson for the initial project proposal, many ideas and discussions throughout the project and initial contributions to the manuscript. We also thank Noah Linden, Ryan Mann, Ashley Montanaro, and Ronald de Wolf for useful discussions and helpful comments on the manuscript.
This work was supported by the National Research Foundation, Singapore and A*STAR under the CQT Bridging Grant and the Quantum Engineering Programme Award number NRF2021-QEP2-02-P05, and by the Bristol Quantum Engineering Centre for Doctoral Training, EPSRC Grant No.\ EP/L015730/1, while at the University of Bristol where most of this project was conducted. This work was carried out using the computational facilities of the Advanced Computing Research Centre, University of Bristol --- http://www.bris.ac.uk/acrc/ --- and the computational facilities of the National University of Singapore.

\bibliography{apssamp}

\providecommand{\noopsort}[1]{}\providecommand{\singleletter}[1]{#1}%
\begin{thebibliography}{30}%
\makeatletter
\providecommand \@ifxundefined [1]{%
 \@ifx{#1\undefined}
}%
\providecommand \@ifnum [1]{%
 \ifnum #1\expandafter \@firstoftwo
 \else \expandafter \@secondoftwo
 \fi
}%
\providecommand \@ifx [1]{%
 \ifx #1\expandafter \@firstoftwo
 \else \expandafter \@secondoftwo
 \fi
}%
\providecommand \natexlab [1]{#1}%
\providecommand \enquote  [1]{``#1''}%
\providecommand \bibnamefont  [1]{#1}%
\providecommand \bibfnamefont [1]{#1}%
\providecommand \citenamefont [1]{#1}%
\providecommand \href@noop [0]{\@secondoftwo}%
\providecommand \href [0]{\begingroup \@sanitize@url \@href}%
\providecommand \@href[1]{\@@startlink{#1}\@@href}%
\providecommand \@@href[1]{\endgroup#1\@@endlink}%
\providecommand \@sanitize@url [0]{\catcode `\\12\catcode `\$12\catcode
  `\&12\catcode `\#12\catcode `\^12\catcode `\_12\catcode `\%12\relax}%
\providecommand \@@startlink[1]{}%
\providecommand \@@endlink[0]{}%
\providecommand \url  [0]{\begingroup\@sanitize@url \@url }%
\providecommand \@url [1]{\endgroup\@href {#1}{\urlprefix }}%
\providecommand \urlprefix  [0]{URL }%
\providecommand \Eprint [0]{\href }%
\providecommand \doibase [0]{https://doi.org/}%
\providecommand \selectlanguage [0]{\@gobble}%
\providecommand \bibinfo  [0]{\@secondoftwo}%
\providecommand \bibfield  [0]{\@secondoftwo}%
\providecommand \translation [1]{[#1]}%
\providecommand \BibitemOpen [0]{}%
\providecommand \bibitemStop [0]{}%
\providecommand \bibitemNoStop [0]{.\EOS\space}%
\providecommand \EOS [0]{\spacefactor3000\relax}%
\providecommand \BibitemShut  [1]{\csname bibitem#1\endcsname}%
\let\auto@bib@innerbib\@empty
\bibitem [{\citenamefont {Kitaev}(2003)}]{kitaev2003fault}%
  \BibitemOpen
  \bibfield  {author} {\bibinfo {author} {\bibfnamefont {A.~Y.}\ \bibnamefont
  {Kitaev}},\ }\bibfield  {title} {\bibinfo {title} {Fault-tolerant quantum
  computation by anyons},\ }\href
  {https://www.sciencedirect.com/science/article/pii/S0003491602000180?casa_token=3F1_4YEMIrMAAAAA:ZSzs4WuIkmBYK390ZAMjYjm0Yl0WInIhfjRuKNDB1dTVDwPiRIXVpMROexi_Osvr2WOo9p0YcQ}
  {\bibfield  {journal} {\bibinfo  {journal} {Annals of Physics}\ }\textbf
  {\bibinfo {volume} {303}},\ \bibinfo {pages} {2} (\bibinfo {year}
  {2003})}\BibitemShut {NoStop}%
\bibitem [{\citenamefont {Dennis}\ \emph {et~al.}(2002)\citenamefont {Dennis},
  \citenamefont {Kitaev}, \citenamefont {Landahl},\ and\ \citenamefont
  {Preskill}}]{dennis2002topological}%
  \BibitemOpen
  \bibfield  {author} {\bibinfo {author} {\bibfnamefont {E.}~\bibnamefont
  {Dennis}}, \bibinfo {author} {\bibfnamefont {A.}~\bibnamefont {Kitaev}},
  \bibinfo {author} {\bibfnamefont {A.}~\bibnamefont {Landahl}},\ and\ \bibinfo
  {author} {\bibfnamefont {J.}~\bibnamefont {Preskill}},\ }\bibfield  {title}
  {\bibinfo {title} {Topological quantum memory},\ }\href
  {https://aip.scitation.org/doi/abs/10.1063/1.1499754} {\bibfield  {journal}
  {\bibinfo  {journal} {Journal of Mathematical Physics}\ }\textbf {\bibinfo
  {volume} {43}},\ \bibinfo {pages} {4452} (\bibinfo {year}
  {2002})}\BibitemShut {NoStop}%
\bibitem [{\citenamefont {Monroe}\ \emph {et~al.}(2014)\citenamefont {Monroe},
  \citenamefont {Raussendorf}, \citenamefont {Ruthven}, \citenamefont {Brown},
  \citenamefont {Maunz}, \citenamefont {Duan},\ and\ \citenamefont
  {Kim}}]{Monroe}%
  \BibitemOpen
  \bibfield  {author} {\bibinfo {author} {\bibfnamefont {C.}~\bibnamefont
  {Monroe}}, \bibinfo {author} {\bibfnamefont {R.}~\bibnamefont {Raussendorf}},
  \bibinfo {author} {\bibfnamefont {A.}~\bibnamefont {Ruthven}}, \bibinfo
  {author} {\bibfnamefont {K.}~\bibnamefont {Brown}}, \bibinfo {author}
  {\bibfnamefont {P.}~\bibnamefont {Maunz}}, \bibinfo {author} {\bibfnamefont
  {L.-M.}\ \bibnamefont {Duan}},\ and\ \bibinfo {author} {\bibfnamefont
  {J.}~\bibnamefont {Kim}},\ }\bibfield  {title} {\bibinfo {title} {Large-scale
  modular quantum-computer architecture with atomic memory and photonic
  interconnects},\ }\href
  {https://journals.aps.org/pra/abstract/10.1103/PhysRevA.89.022317} {\bibfield
   {journal} {\bibinfo  {journal} {Physical Review A}\ }\textbf {\bibinfo
  {volume} {89}},\ \bibinfo {pages} {022317} (\bibinfo {year}
  {2014})}\BibitemShut {NoStop}%
\bibitem [{\citenamefont {Nickerson}(2015)}]{nickerson2015practical}%
  \BibitemOpen
  \bibfield  {author} {\bibinfo {author} {\bibfnamefont {N.}~\bibnamefont
  {Nickerson}},\ }\emph {\bibinfo {title} {Practical fault-tolerant quantum
  computing}},\ \href {https://core.ac.uk/download/pdf/77009079.pdf} {Ph.D.
  thesis},\ \bibinfo  {school} {Imperial College London} (\bibinfo {year}
  {2015})\BibitemShut {NoStop}%
\bibitem [{\citenamefont {Kim}\ \emph {et~al.}(2011)\citenamefont {Kim},
  \citenamefont {Maunz}, \citenamefont {Kim}, \citenamefont {Hussman},
  \citenamefont {Noek}, \citenamefont {Mehta},\ and\ \citenamefont
  {Monroe}}]{kim2011modular}%
  \BibitemOpen
  \bibfield  {author} {\bibinfo {author} {\bibfnamefont {J.}~\bibnamefont
  {Kim}}, \bibinfo {author} {\bibfnamefont {P.}~\bibnamefont {Maunz}}, \bibinfo
  {author} {\bibfnamefont {T.}~\bibnamefont {Kim}}, \bibinfo {author}
  {\bibfnamefont {J.}~\bibnamefont {Hussman}}, \bibinfo {author} {\bibfnamefont
  {R.}~\bibnamefont {Noek}}, \bibinfo {author} {\bibfnamefont {A.}~\bibnamefont
  {Mehta}},\ and\ \bibinfo {author} {\bibfnamefont {C.}~\bibnamefont
  {Monroe}},\ }\bibfield  {title} {\bibinfo {title} {Modular universal scalable
  ion-trap quantum computer ({MUSIQC})},\ }in\ \href
  {https://aip.scitation.org/doi/abs/10.1063/1.3630178} {\emph {\bibinfo
  {booktitle} {AIP Conference Proceedings}}},\ Vol.\ \bibinfo {volume} {1363}\
  (\bibinfo {organization} {American Institute of Physics},\ \bibinfo {year}
  {2011})\ pp.\ \bibinfo {pages} {190--193}\BibitemShut {NoStop}%
\bibitem [{\citenamefont {Knill}\ \emph {et~al.}(2001)\citenamefont {Knill},
  \citenamefont {Laflamme},\ and\ \citenamefont {Milburn}}]{knill2001scheme}%
  \BibitemOpen
  \bibfield  {author} {\bibinfo {author} {\bibfnamefont {E.}~\bibnamefont
  {Knill}}, \bibinfo {author} {\bibfnamefont {R.}~\bibnamefont {Laflamme}},\
  and\ \bibinfo {author} {\bibfnamefont {G.~J.}\ \bibnamefont {Milburn}},\
  }\bibfield  {title} {\bibinfo {title} {A scheme for efficient quantum
  computation with linear optics},\ }\href
  {https://www.nature.com/articles/35051009} {\bibfield  {journal} {\bibinfo
  {journal} {Nature}\ }\textbf {\bibinfo {volume} {409}},\ \bibinfo {pages}
  {46} (\bibinfo {year} {2001})}\BibitemShut {NoStop}%
\bibitem [{\citenamefont {Gimeno-Segovia}\ \emph {et~al.}(2015)\citenamefont
  {Gimeno-Segovia}, \citenamefont {Shadbolt}, \citenamefont {Browne},\ and\
  \citenamefont {Rudolph}}]{Mercedes}%
  \BibitemOpen
  \bibfield  {author} {\bibinfo {author} {\bibfnamefont {M.}~\bibnamefont
  {Gimeno-Segovia}}, \bibinfo {author} {\bibfnamefont {P.}~\bibnamefont
  {Shadbolt}}, \bibinfo {author} {\bibfnamefont {D.~E.}\ \bibnamefont
  {Browne}},\ and\ \bibinfo {author} {\bibfnamefont {T.}~\bibnamefont
  {Rudolph}},\ }\bibfield  {title} {\bibinfo {title} {From three-photon
  {G}reenberger-{H}orne-{Z}eilinger states to ballistic universal quantum
  computation},\ }\href
  {https://journals.aps.org/prl/abstract/10.1103/PhysRevLett.115.020502}
  {\bibfield  {journal} {\bibinfo  {journal} {Physical review letters}\
  }\textbf {\bibinfo {volume} {115}},\ \bibinfo {pages} {020502} (\bibinfo
  {year} {2015})}\BibitemShut {NoStop}%
\bibitem [{\citenamefont {Morley-Short}\ \emph {et~al.}(2019)\citenamefont
  {Morley-Short}, \citenamefont {Gimeno-Segovia}, \citenamefont {Rudolph},\
  and\ \citenamefont {Cable}}]{Morley_Short_2019}%
  \BibitemOpen
  \bibfield  {author} {\bibinfo {author} {\bibfnamefont {S.}~\bibnamefont
  {Morley-Short}}, \bibinfo {author} {\bibfnamefont {M.}~\bibnamefont
  {Gimeno-Segovia}}, \bibinfo {author} {\bibfnamefont {T.}~\bibnamefont
  {Rudolph}},\ and\ \bibinfo {author} {\bibfnamefont {H.}~\bibnamefont
  {Cable}},\ }\bibfield  {title} {\bibinfo {title} {Loss-tolerant teleportation
  on large stabilizer states},\ }\href
  {https://doi.org/10.1088/2058-9565/aaf6c4} {\bibfield  {journal} {\bibinfo
  {journal} {Quantum Science and Technology}\ }\textbf {\bibinfo {volume}
  {4}},\ \bibinfo {pages} {025014} (\bibinfo {year} {2019})}\BibitemShut
  {NoStop}%
\bibitem [{\citenamefont {Morley-Short}\ \emph {et~al.}(2017)\citenamefont
  {Morley-Short}, \citenamefont {Bartolucci}, \citenamefont {Gimeno-Segovia},
  \citenamefont {Shadbolt}, \citenamefont {Cable},\ and\ \citenamefont
  {Rudolph}}]{Morley_Short_2017}%
  \BibitemOpen
  \bibfield  {author} {\bibinfo {author} {\bibfnamefont {S.}~\bibnamefont
  {Morley-Short}}, \bibinfo {author} {\bibfnamefont {S.}~\bibnamefont
  {Bartolucci}}, \bibinfo {author} {\bibfnamefont {M.}~\bibnamefont
  {Gimeno-Segovia}}, \bibinfo {author} {\bibfnamefont {P.}~\bibnamefont
  {Shadbolt}}, \bibinfo {author} {\bibfnamefont {H.}~\bibnamefont {Cable}},\
  and\ \bibinfo {author} {\bibfnamefont {T.}~\bibnamefont {Rudolph}},\
  }\bibfield  {title} {\bibinfo {title} {Physical-depth architectural
  requirements for generating universal photonic cluster states},\ }\href
  {https://doi.org/10.1088/2058-9565/aa913b} {\bibfield  {journal} {\bibinfo
  {journal} {Quantum Science and Technology}\ }\textbf {\bibinfo {volume}
  {3}},\ \bibinfo {pages} {015005} (\bibinfo {year} {2017})}\BibitemShut
  {NoStop}%
\bibitem [{\citenamefont {Edmonds}(1967)}]{edmonds1967optimum}%
  \BibitemOpen
  \bibfield  {author} {\bibinfo {author} {\bibfnamefont {J.}~\bibnamefont
  {Edmonds}},\ }\bibfield  {title} {\bibinfo {title} {Optimum branchings},\
  }\href@noop {} {\bibfield  {journal} {\bibinfo  {journal} {Journal of
  Research of the national Bureau of Standards B}\ }\textbf {\bibinfo {volume}
  {71}},\ \bibinfo {pages} {233} (\bibinfo {year} {1967})}\BibitemShut
  {NoStop}%
\bibitem [{\citenamefont {Kolmogorov}(2009)}]{kolmogorov2009blossom}%
  \BibitemOpen
  \bibfield  {author} {\bibinfo {author} {\bibfnamefont {V.}~\bibnamefont
  {Kolmogorov}},\ }\bibfield  {title} {\bibinfo {title} {Blossom {V}: a new
  implementation of a minimum cost perfect matching algorithm},\ }\href
  {https://link.springer.com/article/10.1007/s12532-009-0002-8} {\bibfield
  {journal} {\bibinfo  {journal} {Mathematical Programming Computation}\
  }\textbf {\bibinfo {volume} {1}},\ \bibinfo {pages} {43} (\bibinfo {year}
  {2009})}\BibitemShut {NoStop}%
\bibitem [{\citenamefont {Davies}\ and\ \citenamefont
  {Parns}(1988)}]{Davies1998}%
  \BibitemOpen
  \bibfield  {author} {\bibinfo {author} {\bibfnamefont {E.~B.}\ \bibnamefont
  {Davies}}\ and\ \bibinfo {author} {\bibfnamefont {L.}~\bibnamefont {Parns}},\
  }\bibfield  {title} {\bibinfo {title} {Trapped modes in acoustic
  waveguides},\ }\href
  {https://academic.oup.com/qjmam/article/51/3/477/1914578} {\bibfield
  {journal} {\bibinfo  {journal} {Q. J. Mech. Appl. Math.}\ }\textbf {\bibinfo
  {volume} {51}},\ \bibinfo {pages} {477} (\bibinfo {year} {1988})}\BibitemShut
  {NoStop}%
\bibitem [{\citenamefont {Wang}\ \emph {et~al.}(2003)\citenamefont {Wang},
  \citenamefont {Harrington},\ and\ \citenamefont
  {Preskill}}]{wang2003confinement}%
  \BibitemOpen
  \bibfield  {author} {\bibinfo {author} {\bibfnamefont {C.}~\bibnamefont
  {Wang}}, \bibinfo {author} {\bibfnamefont {J.}~\bibnamefont {Harrington}},\
  and\ \bibinfo {author} {\bibfnamefont {J.}~\bibnamefont {Preskill}},\
  }\bibfield  {title} {\bibinfo {title} {Confinement-higgs transition in a
  disordered gauge theory and the accuracy threshold for quantum memory},\
  }\href
  {https://www.sciencedirect.com/science/article/pii/S0003491602000192?casa_token=VRO8Z0-n9YUAAAAA:ABRUBIX0wI0j-DX3d9jV9AA0vlWiVHdZlotCLpO4cS_eQSww1HFa0YHYFqc58O2BEVc639xhuw}
  {\bibfield  {journal} {\bibinfo  {journal} {Annals of Physics}\ }\textbf
  {\bibinfo {volume} {303}},\ \bibinfo {pages} {31} (\bibinfo {year}
  {2003})}\BibitemShut {NoStop}%
\bibitem [{\citenamefont {Stace}\ and\ \citenamefont
  {Barrett}(2010)}]{stace2010error}%
  \BibitemOpen
  \bibfield  {author} {\bibinfo {author} {\bibfnamefont {T.~M.}\ \bibnamefont
  {Stace}}\ and\ \bibinfo {author} {\bibfnamefont {S.~D.}\ \bibnamefont
  {Barrett}},\ }\bibfield  {title} {\bibinfo {title} {Error correction and
  degeneracy in surface codes suffering loss},\ }\href
  {https://journals.aps.org/pra/abstract/10.1103/PhysRevA.81.022317} {\bibfield
   {journal} {\bibinfo  {journal} {Physical Review A}\ }\textbf {\bibinfo
  {volume} {81}},\ \bibinfo {pages} {022317} (\bibinfo {year}
  {2010})}\BibitemShut {NoStop}%
\bibitem [{\citenamefont {Duclos-Cianci}\ and\ \citenamefont
  {Poulin}(2010)}]{duclos2010fast}%
  \BibitemOpen
  \bibfield  {author} {\bibinfo {author} {\bibfnamefont {G.}~\bibnamefont
  {Duclos-Cianci}}\ and\ \bibinfo {author} {\bibfnamefont {D.}~\bibnamefont
  {Poulin}},\ }\bibfield  {title} {\bibinfo {title} {Fast decoders for
  topological quantum codes},\ }\href
  {https://journals.aps.org/prl/abstract/10.1103/PhysRevLett.104.050504}
  {\bibfield  {journal} {\bibinfo  {journal} {Physical review letters}\
  }\textbf {\bibinfo {volume} {104}},\ \bibinfo {pages} {050504} (\bibinfo
  {year} {2010})}\BibitemShut {NoStop}%
\bibitem [{\citenamefont {Criger}\ and\ \citenamefont
  {Ashraf}(2018)}]{criger2018multi}%
  \BibitemOpen
  \bibfield  {author} {\bibinfo {author} {\bibfnamefont {B.}~\bibnamefont
  {Criger}}\ and\ \bibinfo {author} {\bibfnamefont {I.}~\bibnamefont
  {Ashraf}},\ }\bibfield  {title} {\bibinfo {title} {Multi-path summation for
  decoding 2{D} topological codes},\ }\href
  {https://quantum-journal.org/papers/q-2018-10-19-102/} {\bibfield  {journal}
  {\bibinfo  {journal} {Quantum}\ }\textbf {\bibinfo {volume} {2}},\ \bibinfo
  {pages} {102} (\bibinfo {year} {2018})}\BibitemShut {NoStop}%
\bibitem [{\citenamefont {Beverland}\ \emph {et~al.}(2019)\citenamefont
  {Beverland}, \citenamefont {Brown}, \citenamefont {Kastoryano},\ and\
  \citenamefont {Marolleau}}]{beverland2019role}%
  \BibitemOpen
  \bibfield  {author} {\bibinfo {author} {\bibfnamefont {M.~E.}\ \bibnamefont
  {Beverland}}, \bibinfo {author} {\bibfnamefont {B.~J.}\ \bibnamefont
  {Brown}}, \bibinfo {author} {\bibfnamefont {M.~J.}\ \bibnamefont
  {Kastoryano}},\ and\ \bibinfo {author} {\bibfnamefont {Q.}~\bibnamefont
  {Marolleau}},\ }\bibfield  {title} {\bibinfo {title} {The role of entropy in
  topological quantum error correction},\ }\href
  {https://iopscience.iop.org/article/10.1088/1742-5468/ab25de/meta?casa_token=AcevT8rfM7UAAAAA:OSpt6R43lYeKyJXEjs_c0KP0qi1Zlr1cY13glSQMZ3bw1BuWqkaTAdedLuBBjRE4z85eYhsM}
  {\bibfield  {journal} {\bibinfo  {journal} {Journal of Statistical Mechanics:
  Theory and Experiment}\ }\textbf {\bibinfo {volume} {2019}},\ \bibinfo
  {pages} {073404} (\bibinfo {year} {2019})}\BibitemShut {NoStop}%
\bibitem [{\citenamefont {Hutter}\ \emph {et~al.}(2014)\citenamefont {Hutter},
  \citenamefont {Wootton},\ and\ \citenamefont {Loss}}]{hutter2014efficient}%
  \BibitemOpen
  \bibfield  {author} {\bibinfo {author} {\bibfnamefont {A.}~\bibnamefont
  {Hutter}}, \bibinfo {author} {\bibfnamefont {J.~R.}\ \bibnamefont
  {Wootton}},\ and\ \bibinfo {author} {\bibfnamefont {D.}~\bibnamefont
  {Loss}},\ }\bibfield  {title} {\bibinfo {title} {Efficient {M}arkov chain
  {M}onte {C}arlo algorithm for the surface code},\ }\href
  {https://journals.aps.org/pra/abstract/10.1103/PhysRevA.89.022326} {\bibfield
   {journal} {\bibinfo  {journal} {Physical Review A}\ }\textbf {\bibinfo
  {volume} {89}},\ \bibinfo {pages} {022326} (\bibinfo {year}
  {2014})}\BibitemShut {NoStop}%
\bibitem [{\citenamefont {Delfosse}\ and\ \citenamefont
  {Nickerson}(2021)}]{delfosse2017almost}%
  \BibitemOpen
  \bibfield  {author} {\bibinfo {author} {\bibfnamefont {N.}~\bibnamefont
  {Delfosse}}\ and\ \bibinfo {author} {\bibfnamefont {N.~H.}\ \bibnamefont
  {Nickerson}},\ }\bibfield  {title} {\bibinfo {title} {Almost-linear time
  decoding algorithm for topological codes},\ }\href
  {https://quantum-journal.org/papers/q-2021-12-02-595/} {\bibfield  {journal}
  {\bibinfo  {journal} {Quantum}\ }\textbf {\bibinfo {volume} {5}},\ \bibinfo
  {pages} {595} (\bibinfo {year} {2021})}\BibitemShut {NoStop}%
\bibitem [{\citenamefont {Cook}\ and\ \citenamefont
  {Rohe}(1999)}]{cook1999computing}%
  \BibitemOpen
  \bibfield  {author} {\bibinfo {author} {\bibfnamefont {W.}~\bibnamefont
  {Cook}}\ and\ \bibinfo {author} {\bibfnamefont {A.}~\bibnamefont {Rohe}},\
  }\bibfield  {title} {\bibinfo {title} {Computing minimum-weight perfect
  matchings},\ }\href
  {https://pubsonline.informs.org/doi/abs/10.1287/ijoc.11.2.138} {\bibfield
  {journal} {\bibinfo  {journal} {INFORMS journal on computing}\ }\textbf
  {\bibinfo {volume} {11}},\ \bibinfo {pages} {138} (\bibinfo {year}
  {1999})}\BibitemShut {NoStop}%
\bibitem [{\citenamefont {Dijkstra}(1959)}]{dijkstra1959note}%
  \BibitemOpen
  \bibfield  {author} {\bibinfo {author} {\bibfnamefont {E.~W.}\ \bibnamefont
  {Dijkstra}},\ }\bibfield  {title} {\bibinfo {title} {A note on two problems
  in connexion with graphs},\ }\href@noop {} {\bibfield  {journal} {\bibinfo
  {journal} {Numerische mathematik}\ }\textbf {\bibinfo {volume} {1}},\
  \bibinfo {pages} {269} (\bibinfo {year} {1959})}\BibitemShut {NoStop}%
\bibitem [{\citenamefont {Yen}(1971)}]{yen1971finding}%
  \BibitemOpen
  \bibfield  {author} {\bibinfo {author} {\bibfnamefont {J.~Y.}\ \bibnamefont
  {Yen}},\ }\bibfield  {title} {\bibinfo {title} {Finding the k shortest
  loopless paths in a network},\ }\href
  {https://pubsonline.informs.org/doi/abs/10.1287/mnsc.17.11.712} {\bibfield
  {journal} {\bibinfo  {journal} {management Science}\ }\textbf {\bibinfo
  {volume} {17}},\ \bibinfo {pages} {712} (\bibinfo {year} {1971})}\BibitemShut
  {NoStop}%
\bibitem [{\citenamefont {Nickerson}\ \emph {et~al.}(2013)\citenamefont
  {Nickerson}, \citenamefont {Li},\ and\ \citenamefont
  {Benjamin}}]{nickerson2013topological}%
  \BibitemOpen
  \bibfield  {author} {\bibinfo {author} {\bibfnamefont {N.~H.}\ \bibnamefont
  {Nickerson}}, \bibinfo {author} {\bibfnamefont {Y.}~\bibnamefont {Li}},\ and\
  \bibinfo {author} {\bibfnamefont {S.~C.}\ \bibnamefont {Benjamin}},\
  }\bibfield  {title} {\bibinfo {title} {Topological quantum computing with a
  very noisy network and local error rates approaching one percent},\ }\href
  {https://www.nature.com/articles/ncomms2773} {\bibfield  {journal} {\bibinfo
  {journal} {Nature communications}\ }\textbf {\bibinfo {volume} {4}},\
  \bibinfo {pages} {1} (\bibinfo {year} {2013})}\BibitemShut {NoStop}%
\bibitem [{\citenamefont {Fowler}\ \emph {et~al.}(2012)\citenamefont {Fowler},
  \citenamefont {Whiteside}, \citenamefont {McInnes},\ and\ \citenamefont
  {Rabbani}}]{fowler2012topological}%
  \BibitemOpen
  \bibfield  {author} {\bibinfo {author} {\bibfnamefont {A.~G.}\ \bibnamefont
  {Fowler}}, \bibinfo {author} {\bibfnamefont {A.~C.}\ \bibnamefont
  {Whiteside}}, \bibinfo {author} {\bibfnamefont {A.~L.}\ \bibnamefont
  {McInnes}},\ and\ \bibinfo {author} {\bibfnamefont {A.}~\bibnamefont
  {Rabbani}},\ }\bibfield  {title} {\bibinfo {title} {Topological code
  autotune},\ }\href
  {https://journals.aps.org/prx/abstract/10.1103/PhysRevX.2.041003} {\bibfield
  {journal} {\bibinfo  {journal} {Physical Review X}\ }\textbf {\bibinfo
  {volume} {2}},\ \bibinfo {pages} {041003} (\bibinfo {year}
  {2012})}\BibitemShut {NoStop}%
\bibitem [{dor()}]{doriguello_data}%
  \BibitemOpen
  \href {https://github.com/joaodoriguello/Asynchronous-decoding} {\bibinfo
  {title}
  {https://github.com/joaodoriguello/asynchronous-decoding}}\BibitemShut
  {NoStop}%
\bibitem [{\citenamefont {Watson}\ \emph {et~al.}(2015)\citenamefont {Watson},
  \citenamefont {Anwar},\ and\ \citenamefont {Browne}}]{watson2015fast}%
  \BibitemOpen
  \bibfield  {author} {\bibinfo {author} {\bibfnamefont {F.~H.}\ \bibnamefont
  {Watson}}, \bibinfo {author} {\bibfnamefont {H.}~\bibnamefont {Anwar}},\ and\
  \bibinfo {author} {\bibfnamefont {D.~E.}\ \bibnamefont {Browne}},\ }\bibfield
   {title} {\bibinfo {title} {Fast fault-tolerant decoder for qubit and qudit
  surface codes},\ }\href
  {https://journals.aps.org/pra/abstract/10.1103/PhysRevA.92.032309} {\bibfield
   {journal} {\bibinfo  {journal} {Physical Review A}\ }\textbf {\bibinfo
  {volume} {92}},\ \bibinfo {pages} {032309} (\bibinfo {year}
  {2015})}\BibitemShut {NoStop}%
\bibitem [{\citenamefont {Ohno}\ \emph {et~al.}(2004)\citenamefont {Ohno},
  \citenamefont {Arakawa}, \citenamefont {Ichinose},\ and\ \citenamefont
  {Matsui}}]{ohno2004phase}%
  \BibitemOpen
  \bibfield  {author} {\bibinfo {author} {\bibfnamefont {T.}~\bibnamefont
  {Ohno}}, \bibinfo {author} {\bibfnamefont {G.}~\bibnamefont {Arakawa}},
  \bibinfo {author} {\bibfnamefont {I.}~\bibnamefont {Ichinose}},\ and\
  \bibinfo {author} {\bibfnamefont {T.}~\bibnamefont {Matsui}},\ }\bibfield
  {title} {\bibinfo {title} {Phase structure of the random-plaquette {Z}2 gauge
  model: accuracy threshold for a toric quantum memory},\ }\href
  {https://www.sciencedirect.com/science/article/pii/S055032130400481X?casa_token=1lTt6QtuVsYAAAAA:GT6ihUvtB9kZeUga24u4EMHkXEWjEXzgI0k0aK127uu3sPrbREchn-nUfRLSZ3lYqGJLayxZkQ}
  {\bibfield  {journal} {\bibinfo  {journal} {Nuclear physics B}\ }\textbf
  {\bibinfo {volume} {697}},\ \bibinfo {pages} {462} (\bibinfo {year}
  {2004})}\BibitemShut {NoStop}%
\bibitem [{\citenamefont {Stephens}(2014)}]{stephens2014fault}%
  \BibitemOpen
  \bibfield  {author} {\bibinfo {author} {\bibfnamefont {A.~M.}\ \bibnamefont
  {Stephens}},\ }\bibfield  {title} {\bibinfo {title} {Fault-tolerant
  thresholds for quantum error correction with the surface code},\ }\href
  {https://journals.aps.org/pra/abstract/10.1103/PhysRevA.89.022321} {\bibfield
   {journal} {\bibinfo  {journal} {Physical Review A}\ }\textbf {\bibinfo
  {volume} {89}},\ \bibinfo {pages} {022321} (\bibinfo {year}
  {2014})}\BibitemShut {NoStop}%
\bibitem [{\citenamefont {Harrington}(2004)}]{harrington2004analysis}%
  \BibitemOpen
  \bibfield  {author} {\bibinfo {author} {\bibfnamefont {J.~W.}\ \bibnamefont
  {Harrington}},\ }\emph {\bibinfo {title} {Analysis of quantum
  error-correcting codes: symplectic lattice codes and toric codes}},\ \href
  {https://thesis.library.caltech.edu/1747/1/jimh_thesis.pdf} {Ph.D. thesis},\
  \bibinfo  {school} {California Institute of Technology} (\bibinfo {year}
  {2004})\BibitemShut {NoStop}%
\bibitem [{\citenamefont {Fredman}\ and\ \citenamefont
  {Tarjan}(1987)}]{fredman1987fibonacci}%
  \BibitemOpen
  \bibfield  {author} {\bibinfo {author} {\bibfnamefont {M.~L.}\ \bibnamefont
  {Fredman}}\ and\ \bibinfo {author} {\bibfnamefont {R.~E.}\ \bibnamefont
  {Tarjan}},\ }\bibfield  {title} {\bibinfo {title} {Fibonacci heaps and their
  uses in improved network optimization algorithms},\ }\href
  {https://dl.acm.org/doi/abs/10.1145/28869.28874} {\bibfield  {journal}
  {\bibinfo  {journal} {Journal of the ACM (JACM)}\ }\textbf {\bibinfo {volume}
  {34}},\ \bibinfo {pages} {596} (\bibinfo {year} {1987})}\BibitemShut
  {NoStop}%
\end{thebibliography}%


\providecommand{\noopsort}[1]{}\providecommand{\singleletter}[1]{#1}%
%

\appendix

\section{Simulation methods}
\label{app:app_sim}

The simulations were all carried out in an $L\times L$ periodic lattice with $L\in\{10,12,14\}$ and repeated a number of $10^5$ to $10^6$ times. The simulations used C++ and were carried out using the computational facilities of the Advanced Computing Research Centre, University of Bristol, and the computational facilities of the National University of Singapore. The nodes had 16 and 64 cores. Each single threshold value was computed by plotting the decoding success probability for a range of $16$ values of the physical error $p$, and took between $15$ hours for our simple decoders (BG and AP decoders) to 48 hours for our more complex decoders (CG decoder). All codes and raw data can be found in~\cite{doriguello_data}.

\subsection{Discrete measurement}

We perform $N_s$ measurement rounds and generate a simple syndrome graph of size $L\times L \times N_s$ with $N_s = \lceil 2/s \rfloor L$, where $\lceil x \rfloor$ denotes the closest integer to $x$. We assume periodic boundaries in space, and open boundaries in time, corresponding to initialisation and destructive measurement of a toric code. The last measurement round is taken to be perfect in order to guarantee the existence of a perfect matching of the anyons. 

At each measurement round we flip the value of each qubit with probability $p_\Delta$, the simulation error, after which we perform the stabiliser measurements, each with probability $s$. The stabiliser outcomes, if successful, are flipped with probability $q$. The time scale is defined such that a stabiliser outcome per qubit is obtained at an average rate of $1$, i.e., after $1/s$ measurement rounds, meaning that consecutive measurement rounds are separated by a time interval of $s$. The physical error $p$ is related to $p_\Delta$ according to Eq.~\eqref{eq:eq2.1}. We always fix $q=p$.

The resulting simple syndrome graph with error configuration is then processed to construct the matching graph. Depending on the decoder this may first involve performing edge contraction of erased edges.

\subsection{Continuous measurement}

In the limit $s\to 0$ we cannot sample discretely and instead generate the contracted syndrome graph directly. We first note that the number of bit-flips that a qubit suffers in the discrete measurement regime is a Binomial distribution $B(N_s,p_\Delta)$. Thus, in the limit $s\to 0$ and $N_s\to\infty$, such Binomial distribution converges towards a \emph{Poisson distribution} with parameter $\lim_{s\to 0} N_s \cdot p_\Delta$ according to the \emph{Poisson limit theorem}. Using Eq.~\eqref{eq:eq2.1b} and that $N_s = T/s$, where $T$ is time corresponding to the last measurement round, we obtain
\begin{align*}
    \lim_{s\to 0} N_s \cdot p_\Delta = \frac{T}{2}\lim_{s\to 0} \frac{1 - (1 - 2p)^{s}}{s} = \frac{T}{2}\ln\left(\frac{1}{1 - 2p}\right).
\end{align*}
A similar reasoning applies to stabiliser measurements: the number of successful stabiliser measurements in the limit $s\to 0$ has a Poisson distribution with parameter $\lim_{s\to 0} s\cdot N_s = T$. The simulation for the continuous asynchronous regime ($s=0$) is then performed by first setting a time interval $T=2L$ and a physical error $p$ and then, for each qubit, sampling the number of bit-flips it suffers from a Poisson distribution with parameter $\frac{T}{2} \ln(1/(1-2p))$ and distributing these bit-flips uniformly at random along the time interval $(0,T)$. The same is done with the measurements: for each stabiliser operator, we sample a number of successful faulty measurements from a Poisson distribution with parameter $T$ and uniformly distribute these measurements along the time interval $(0,T)$. We also include perfect measurements at time $T$ to guarantee the existence of a perfect matching.

\subsection{Estimating the $P_E$ terms}

We also computed the average relative size between the first few $P_E$ terms and $P_0$ from Eq.~\eqref{eq:Pij_iid}. The ratios were obtained by averaging over \emph{matched} pairs of anyons by the CG decoder in a typical simulation as follows: given an $L\times L$ lattice, we first set a synchronicity $s$, a physical error $p$ and a measurement error $q=p$. The value of $p$ was chosen as the threshold of the CG decoder from Fig.~\ref{fig:fig4a} at the given $s$. The number of measurement rounds was set as $N_s = \left\lceil 2/s\right\rfloor L$ for $s\in(0,1]$, and time interval $T=2L$ for $s=0$. 
We then applied the usual procedure just described above of introducing physical errors and measuring the stabilisers to obtain a random contracted syndrome graph and a set of anyons. The ratio $P_E/P_0$ is averaged over all the anyon pairs that would be matched by the CG decoder. We stress that we do not average over all anyon pairs, since anyons far away are disregarded by the decoder. The final result was averaged over other random contracted syndrome graphs. The number of samples over random contracted syndrome graphs was set to between $1000$ and $25000$, depending on the lattice size. 
Computing $P_E$ for each pair of anyons was performed via Yen's algorithm~\cite{yen1971finding}, which is a generalisation of Dijkstra's algorithm for computing the $k$-shortest loopless paths in a graph with non-negative edge cost.\footnote{If the length of the $E$-shortest path in the contracted syndrome graph is $l_E = \sum_e \ln((1-p_e)/p_e)$, then $P_E \propto \exp(-l_E)$.}

\section{Extra Results}
\label{app:appA0}

\subsection{Fully synchronous i.i.d.\ error model}

In the fully synchronous ($s=1$) regime with faulty measurements and i.i.d.\ error model, the first-order degeneracy term $\Omega_{0}$ from Eq.~\eqref{eq:eq_f2} between two anyons $i$ and $j$ with coordinates $(x_i,y_i,t_i)$ and $(x_j,y_j,t_j)$ can be explicitly calculated as
\begin{align}
    \label{eq:iid-count}
    \Omega_{0}(i,j) = \frac{(\Delta (x_i,x_j) + \Delta (y_i,y_j) + |t_i - t_j|)!}{(\Delta (x_i,x_j))!(\Delta (y_i,y_j))!(|t_i - t_j|)!},
\end{align}
which is the number of ways one can take $\Delta (x_i,x_j)$ steps in the $x$-direction, $\Delta (y_i,y_j)$ steps in the $y$-direction and $|t_i - t_j|$ steps in the $t$-direction. The above expression was previously considered by~\cite{stace2010error,watson2015fast,beverland2019role}.

\begin{figure}[t]
    \centering
    \includegraphics[width=\columnwidth]{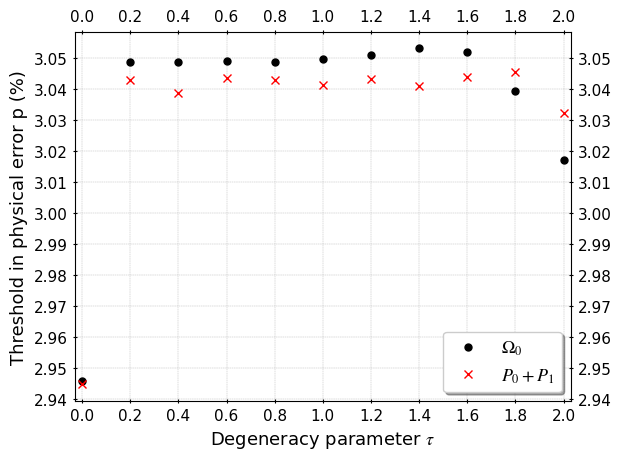}
    \caption{Threshold dependence with the degeneracy parameter $\tau$ for the MWPM decoder with either the first-order degeneracy term $\Omega_0$ or the two highest probability terms $P_0$ and $P_1$ in the fully synchronous ($s=1$) regime with faulty measurements and i.i.d.\ error model. The thresholds at $\tau=0$ and $\tau=1$ are $2.937\%\pm 0.002\%$ and $3.050\%\pm 0.002\%$, respectively.}
    \label{fig:3D_tau}
\end{figure}

The first-order degeneracy term can be included into the MWPM decoder through the weight assignment 
\begin{align}
    \label{eq:iid-weight}
    w_{ij} = l_0(i,j)\ln\left(\frac{1-p}{p}\right) - \tau\ln{\Omega_{0}(i,j)},
\end{align}
similarly to Eq.~\eqref{eq:eq6.C1}, where $\tau$ is a tunable degeneracy parameter. Fig.~\ref{fig:3D_tau} shows the effect of the first-order degeneracy term $\Omega_{0}$ on the threshold values using the MWPM decoder as a function of $\tau$. The threshold value at $\tau = 0$ corresponds to the one of the usual MPMW decoder for the toric code, and was computed to be $2.937\%\pm 0.002\%$, in agreement with past results~\cite{wang2003confinement,ohno2004phase,stephens2014fault,nickerson2015practical,harrington2004analysis}. Moreover, the dependence on $\tau$ is very similar to the one observed in~\cite[Figure~9]{stace2010error} with perfect stabiliser measurements. In particular, the threshold does not peak at $\tau=1$, as one would expect, but around $\tau=1.4$, where the matching algorithm favours more degenerate paths, and by $\tau=2$ it has already dropped significantly. The threshold at $\tau=1$ is $3.050\%\pm 0.002\%$.

Furthermore, Fig.~\ref{fig:3D_tau} also shows the effect of considering only the second highest term $P_1$ from Eq.~\eqref{eq:Pij_iid} on top of $P_0$, instead of the first $\Omega_0$ highest terms (replace $\Omega_0$ with $1+P_1/P_0$ in Eq.~\eqref{eq:iid-weight}). We see that most of the first-order-degeneracy-term improvement on the threshold comes from $P_1$ alone, and is a consequence of Fig.~\ref{fig:fig9}.

\subsection{Average Position decoder}

\begin{figure}[ht]
    \centering
    \includegraphics[width=\columnwidth]{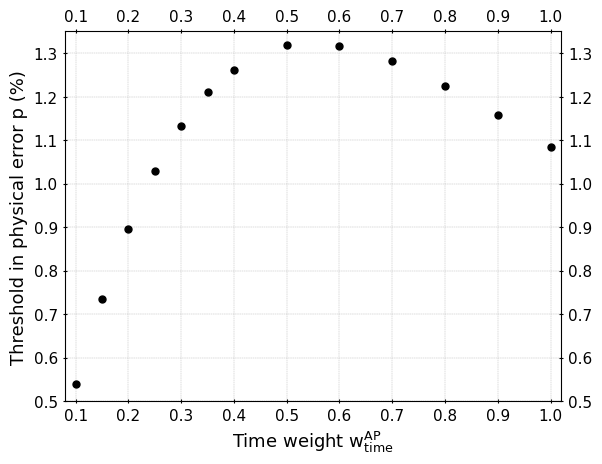}
    \caption{Threshold dependence of the AP decoder with the time weight $w_{\rm time}^{\rm AP}$ in the continuous asynchronous ($s=0$) regime with faulty measurements.}
    \label{fig:figA1}
\end{figure}

\begin{figure}[h!]
    \centering
    \includegraphics[width=\columnwidth]{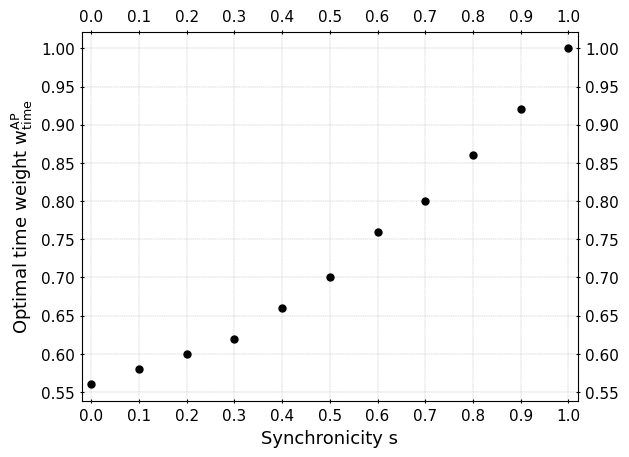}
    \caption{Optimal time weight $w_{\rm time}^{\rm AP}$ versus synchronicity $s$.}
    \label{fig:figA2}
\end{figure}

The AP decoder from Sec.~\ref{sec:sec9} was optimised in terms of the time parameter $w_{\rm time}^{\rm AP}$ (see Eq.~\eqref{eq:eq1}). We explore its dependence on $w_{\rm time}^{\rm AP}$ in Figs.~\ref{fig:figA1} and~\ref{fig:figA2}. More specifically, Fig.~\ref{fig:figA1} shows the thresholds dependence on $w_{\rm time}^{\rm AP}$ in the continuous asynchronous regime ($s=0$), while Fig.~\ref{fig:figA2} shows the optimal value of the $w_{\rm time}^{\rm AP}$ parameter at which the threshold is maximum as a function of $s$.

The overall shape of Fig.~\ref{fig:figA1} is expected: both underestimation (large $w_{\rm time}^{\rm AP}$) and overestimation (small $w_{\rm time}^{\rm AP}$) of the time weight between anyons worsen the performance of the AP decoder. 
An optimal time weight $w_{\rm time}^{\rm AP}$ should be observed. Such point, in Fig.~\ref{fig:figA1} at $w_{\rm time}^{\rm AP} \approx 0.56$, suggests that placing anyons in the middle of parity blocks ``shortens'' the time distance, although we currently do not have deeper arguments explaining this fact. Regarding Fig.~\ref{fig:figA2}, we expect a smooth interpolation between $w_{\rm time}^{\rm AP} \approx 0.56$ at $s=0$ and $w_{\rm time}^{\rm AP}=1$ at $s=1$.


\subsection{Block Graph decoder}

\subsubsection{Time weight parameter}

\begin{figure}[ht]
    \centering
    \includegraphics[width=\columnwidth]{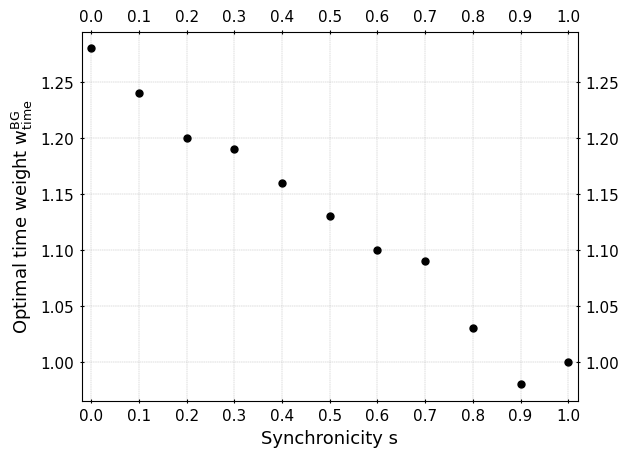}
    \caption{Optimal time weight $w_{\rm time}^{\rm BG}$ versus synchronicity $s$.}
    \label{fig:BG_time_dependence}
\end{figure}

Similarly to the AP decoder, the BG decoder from Sec.~\ref{sec:sec9} was optimised in terms of the time parameter $w_{\rm time}^{\rm BG}$ (see Eq.~\eqref{eq:app-simple}). We explore its dependence on $w_{\rm time}^{\rm BG}$ in Fig.~\ref{fig:BG_time_dependence}. At $s=0$, we observe an optimal time weight $w^{\rm BG}_{\rm time} \approx 1.28$, which points to the fact that, in opposition to the AP decoder, the BG decoder ``stretches'' the time distance. Moreover, during the simulations we noticed that the threshold of the BG decoder is quite insensitive to changes on $w^{\rm BG}_{\rm time}$. This explains the apparent non-smoothness interpolation from $s=0$ to $s=1$: simulation inaccuracies hide the exact optimal point in a relatively large interval of possible values.

\subsubsection{Degeneracy}

\begin{figure}[t]
    \centering
    \includegraphics[trim={4.5cm 17cm 5cm 4.5cm},clip,width=0.6\textwidth]{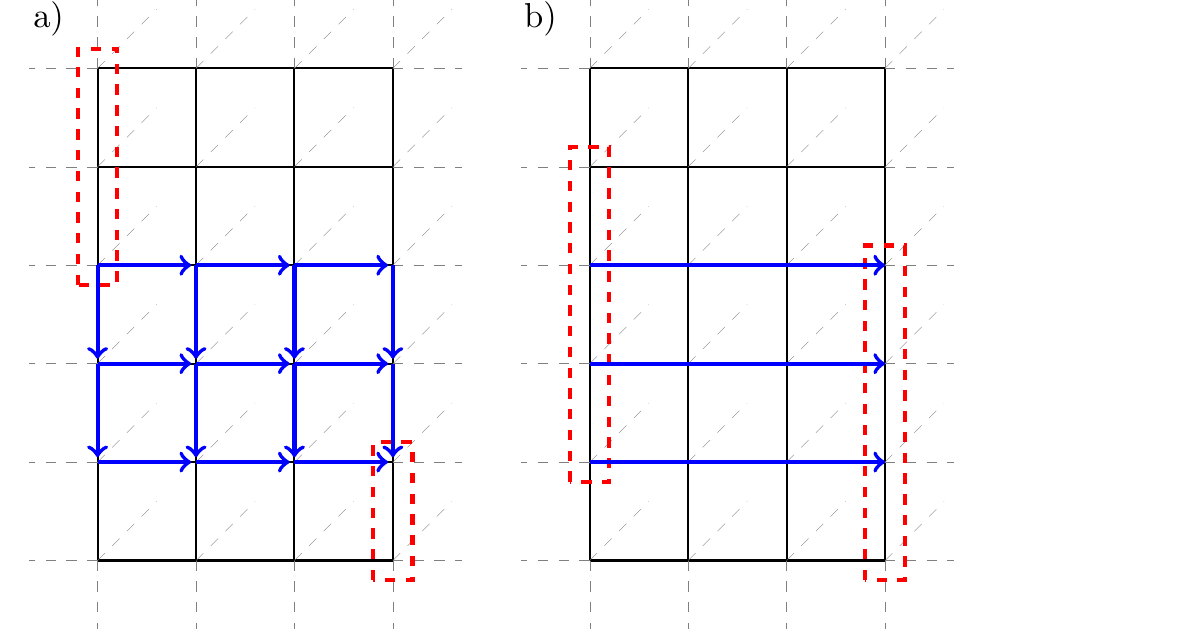}
    \caption{Number $\Omega_0$ of error chains (shortest paths) between two anyon blocks. (a) If the anyon blocks do not overlap in time, $\Omega_0$ is the number of paths between their closest points. In the example there are $(3+2)!/(2!3{!}) = 10$ shortest paths. (b) If the anyon blocks overlap in time, $\Omega_0$ is proportional to their time overlap.}
    \label{fig:simple_deg}
\end{figure}

The BG decoder uses the weight assignment from Eq.~\eqref{eq:app-simple}. The net effect is to disregard erased edges between two anyon blocks and view a pair of anyon blocks as embedded in a cubic syndrome graph: space-like edges have an associated $p_\Delta$ error, while time-like edges have an associated $p\cdot s \approx p_\Delta$ error (since a measurement error occurs with probability $q=p$ if the stabiliser measurement is successful). The scenario is thus similar to the synchronous regime with i.i.d.\ errors and we can introduce high-order degeneracy terms like $\Omega_0$ by counting paths similarly to what was done in the previous section. Consider two anyon blocks $(x_i,y_i,t_{i1},t_{i2})$ and $(x_j,y_j,t_{j1},t_{j2})$. If they do not overlap in time, then $\Omega_0$ is the number of shortest paths between their closest points, given by Eq.~\eqref{eq:iid-count} with $t_{i1}/s$ and $t_{j2}/s$ if $t_{i1}>t_{j2}$ and with $t_{i2}/s$ and $t_{j1}/s$ if $t_{j1}>t_{i2}$ (see Fig.~\ref{fig:simple_deg}). If the blocks overlap in time, then
\begin{multline*}
    \Omega_0(i,j) = \left(\min\left(\frac{t_{i2}}{s},\frac{t_{j2}}{s}\right) - \max\left(\frac{t_{i1}}{s},\frac{t_{j1}}{s}\right)\right)\\
    \times\frac{(\Delta(x_i,x_j) + \Delta(y_i,y_j))!}{(\Delta(x_i,x_j))!(\Delta(y_i,y_j))!},
\end{multline*}
which is the number of paths with zero time-like edges connecting both blocks (remember that the time scale is such that measurement rounds happen at times proportional to $s$). Hence, similarly to Eq.~\eqref{eq:iid-weight}, degeneracy can be introduced into the BG decoder via the weighting assignment
\begin{multline*}
    w_{ij} = w_{\rm time}^{\rm BG}\max(t_{i1}-t_{j2},t_{j1}-t_{i2},0)\\
    + \Delta(x_i,x_j) + \Delta(y_i,y_j) - \tau \ln\Omega_0(i,j)/\ln\left(\frac{1-p_\Delta}{p_\Delta}\right),
\end{multline*}
where again $\tau$ is a tunable degeneracy parameter.

\begin{figure}[ht]
    \centering
    \includegraphics[width=\columnwidth]{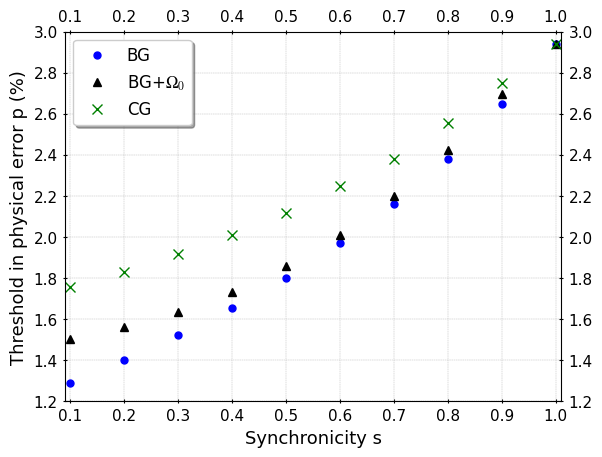}
    \caption{Threshold comparison between the CG decoder and the BG decoder with and without the first-order degeneracy term $\Omega_0$. The threshold of the BG decoder with degeneracy was optimised in terms of the degeneracy parameter $\tau$.}
    \label{fig:approx-simple}
\end{figure}

In Fig.~\ref{fig:approx-simple} we compare the thresholds of the BG decoder enhanced by the degeneracy term $\Omega_0$ with its base case and also with the CG decoder. We can see that degeneracy can improve the BG decoder's performance, specially for high asynchronism, albeit not sufficiently to reach CG decoder's level. Moreover, such improvement by degeneracy does not apply in the limit $s\to 0$, since the simple syndrome graph is not well defined and the naive method of counting paths is not possible anymore, therefore we limit the horizontal scale in Fig.~\ref{fig:approx-simple} to $s\in[0.1,1]$.

\subsection{Degeneracy in the Contracted Graph decoder}

\begin{figure}[ht]
    \centering
    \includegraphics[width=\columnwidth]{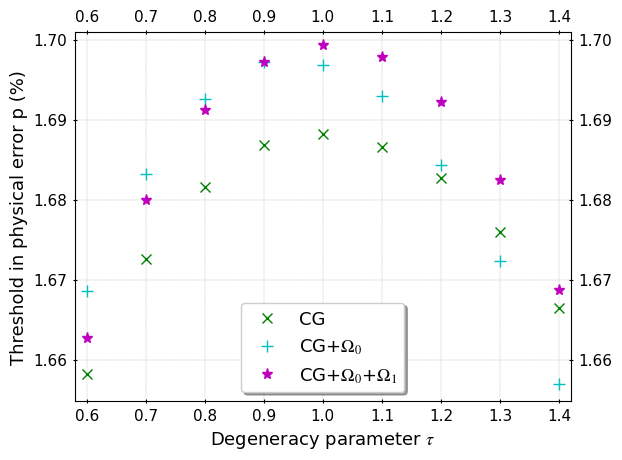}
    \caption{Threshold comparison between the CG decoder and its degeneracy versions with $\Omega_{0}$ and with both $\Omega_{0}$ and $\Omega_{1}$. The comparison is with respect to the degeneracy parameter $\tau$ in the continuous asynchronous regime ($s=0$).}
    \label{fig:figA3}
\end{figure}

The improvement provided by the degeneracy terms $\Omega_{0}$ and $\Omega_{1}$ is quite small, as evident by Fig.~\ref{fig:fig4a}. We explore such improvement in more details for the continuous asynchronous regime. As mentioned in Eq.~(\ref{eq:eq6.C1}), the weight assignment in the CG decoder with degeneracy is $l_0\ln{p^{-1}} - \tau\ln\big(\Omega_{0} + p\cdot\Omega_{1}\big)$. The usual CG decoder without degeneracy can be approximated by using $\tau\ln\delta_{0}$ instead of $\tau\ln\left(\Omega_{0} + p\cdot\Omega_{1}\right)$, where $\delta_0 = \prod_{e\in E_0}\omega(e)$ for the most likely error chain $E_0$. Indeed, from Eq.~\eqref{eq:6.3approximation2}, the CG decoder optimises $\max_E\ln(p^{|E|}\delta_E) = -\min_E\big[|E|\ln{p^{-1}}-\ln{\delta_E}\big]$. We can also define an intermediate instance where $\tau\ln\Omega_{0}$ is used instead of the full $\tau\ln\left(\Omega_{0} + p\cdot\Omega_{1}\right)$. Thus the following weight assignments are possible:
\begin{align*}
    w_{ij} = \begin{cases}
        l_0(i,j)\ln{p^{-1}} - \tau\ln\delta_{0},\\
        l_0(i,j)\ln{p^{-1}} - \tau\ln\Omega_{0},\\
        l_0(i,j)\ln{p^{-1}} - \tau\ln\big(\Omega_{0} + p\cdot\Omega_{1}\big).
    \end{cases}
\end{align*}
In Fig.~\ref{fig:figA3} we depict the improvement provided by including only the degeneracy term $\Omega_{0}$, and by including both $\Omega_{0}$ and $\Omega_{1}$. As expected, the introduction of more degeneracy improves the CG decoder threshold values. Moreover, the decoders' performances peak around $\tau = 1$, as expected given the discussion leading to Eq.~\eqref{eq:eq6B.2}: the probability $P_0$ considered by the CG decoder is best approximated by the case $\tau=1$.

\subsection{$P_E$ terms}

\begin{figure*}[t]
 \begin{center}
    \subfigure[$\langle P_1/P_0\rangle$]{\label{fig:fig8a}\includegraphics[width=0.49\textwidth]{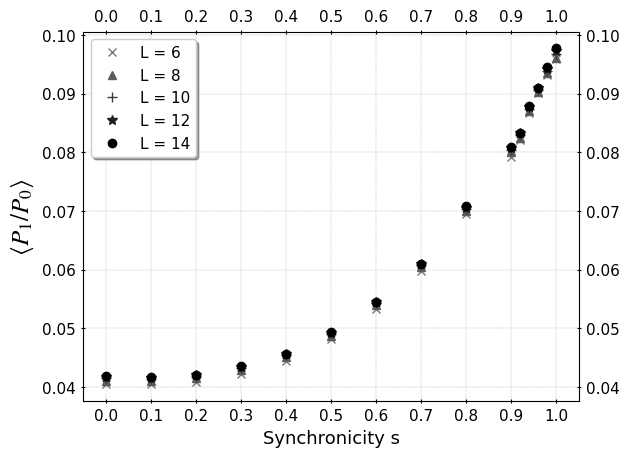}}
    \subfigure[$\langle P_2/P_0\rangle$]{\label{fig:fig8b}\includegraphics[width=0.49\textwidth]{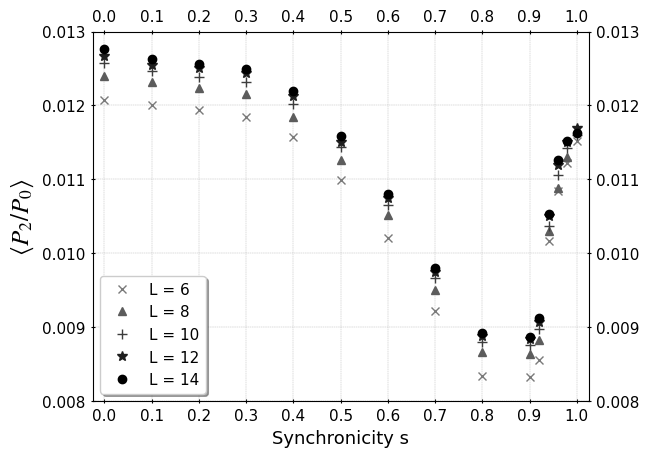}}
    \end{center}
    \caption{Average ratios $\langle P_1/P_0\rangle$ and $\langle P_2/P_0\rangle$ as a function of synchronicity $s$ for different sizes $L$ of an $L\times L\times N_s$ lattice. Here $N_s = \left\lceil 2/s\right\rfloor L$ for $s\in(0,1]$ and $N_0=T=2L$ for $s=0$. The ratios were averaged over random contracted syndrome graphs given an $L\times L\times N_s$ lattice, a synchronicity $s$, a physical error $p$ and a measurement error $q=p$. Given a synchronicity $s$, the value of $p$ was chosen as the threshold of the CG decoder at $s$ from Fig.~\ref{fig:fig4a}.}\label{fig:fig8}
\end{figure*}

Fig.~\ref{fig:fig8} numerically explores the average ratio between some of the first $P_E$ terms and $P_0$ for different sizes of the toric code. More specifically, Fig.~\ref{fig:fig8a} computes the average $\langle P_1/P_0\rangle$ as a function of the synchronicity $s$ in an $L\times L$ lattice for different sizes $L$. Fig.~\ref{fig:fig8b} shows $\langle P_2/P_0\rangle$ in a similar fashion. Turning to the results, in Fig.~\ref{fig:fig8} we see that $P_1$ is between $4\%$-$10\%$ of $P_0$ on average, while $P_2$ corresponds to only $1\%$ of $P_0$ on average. The contribution of $P_1$ diminishes fairly rapidly with $s$, thus explaining the abrupt advantage drop from degeneracy observed in Fig.~\ref{fig:fig4} between $s=1$ and $s=0.9$. Moreover, Fig.~\ref{fig:fig8b} shows an interesting effect which we do not have an explanation for: $P_2$ has a sudden decrease with synchronicity, which further lessens the role of degeneracy in the CG decoder when moving away from $s=1$, followed by a latter improvement with more asynchronism, albeit not enough to make up for $P_1$.

\section{Dijkstra's algorithm}
\label{app:appA}

The weight between two anyons in the CG decoder is obtained via the metric $d_C$ (Eq.~\eqref{eq:eq6B.2}), which means that we are required to calculate shortest paths between two vertices in a graph. In order to do so efficiently, we used Dijkstra's algorithm~\cite{dijkstra1959note}. Its run time is $O(|V|^2)$, where $|V|$ is the number of vertices in the graph. It is possible to improve its complexity to $O(|E| + |V|\log{|V|})$, where $|E|$ is the number of edges, by replacing the min-priority queue from the original algorithm by a Fibonacci heap min-priority queue~\cite{fredman1987fibonacci}. Here we used a binary min-priority heap, which is a heap data structure that takes the form of a binary tree, and a common variant of Dijkstra's algorithm that fixes a source vertex and calculates the shortest paths from it to all the other vertices in the graph, thus producing a shortest-path tree.

The algorithm works by initialising two values:
\begin{itemize}
    \item \texttt{dist[]}, an array of distances from the \texttt{source} vertex to each vertex in the syndrome graph $S$. Initially, $\texttt{dist[source]} = 0$ and $\texttt{dist[}v\texttt{]} = \infty$ for all the other vertices $v$. As the algorithm progresses, the distance from \texttt{source} to each vertex is updated.
    \item $Q$, a min-priority queue of unvisited vertices. Initially $Q$ contains all the vertices. At the end of the algorithm, $Q$ is empty.
\end{itemize}
A min-priority queue is an abstract data type that provides $3$ basic operations: \texttt{add\_with\_priority()}, \texttt{decrease\_priority()} and \texttt{extract\_min()}. The operation \texttt{add\_with\_priority(}$v$, \texttt{dist[}$v$\texttt{])} adds $v$ based on the value \texttt{dist[}$v$\texttt{]}. A min-priority queue will order its vertices $v$ based on the increasing value of \texttt{dist[}$v$\texttt{]}. The \texttt{decrease\_priority(}$v$\texttt{, dist[}$v$\texttt{])} updates the ordering according to a new value \texttt{dist[}$v$\texttt{]} of vertex $v$. And the operation \texttt{extract\_min()} extracts the vertex with the minimum distance (located at the root of a binary or Fibonacci min-priority heap).


In summary, the algorithm first initialises the array of distances and the min-priority queue. Then, at each step, the vertex $u$ with minimum \texttt{dist[}$u$\texttt{]} is extracted from $Q$ and set as the \texttt{current} vertex. We consider all its children and calculate their tentative distances through \texttt{current}, i.e., $\texttt{dist[current]} + d_S(\texttt{current},\texttt{child})$. If \texttt{dist[child]} is greater than this tentative distance, then the distance of the child vertex is updated to the tentative value. We repeat this process until $Q$ is empty.

We provide a pseudo-code for Dijkstra's algorithm.

\begin{algorithm}[H]
    \caption{Dijkstra's algorithm}
    \KwInput{Syndrome graph $C$ and \texttt{source} vertex}
    create min-priority queue $Q$\;
    \ForAll{\upshape vertex $v\in C$} {
        \texttt{dist[$v$]} $\gets$ INFINITY\;
        $Q$.\texttt{add\_with\_priority($v$, dist[$v$])}\;}
    \texttt{dist[source]} $\gets 0$\;
    Q.\texttt{decrease\_priority(source, dist[source])}\;
    \While{$Q \neq \emptyset$} {
        $u \gets Q$.\texttt{extract\_min()}\;
        \For{\upshape each neighbor $v$ of $u$} {
            aux $\gets$ \texttt{dist[$u$]} + $d_C(u,v)$\;
            \uIf{\upshape aux $<$ \texttt{dist[$v$]}} {
                \texttt{dist[$v$]} $\gets$ aux\;
                Q.\texttt{decrease\_priority($v$, dist[$v$])}\;}
        }
    }
    \Return \texttt{dist[]}\;
\end{algorithm}

\subsection{Degeneracy terms}

It is possible to use Dijkstra's algorithm in order to calculate the degeneracies $\Omega_{0}$ and $\Omega_{1}$ between the source vertex and all other vertices. This is done by using a simplified version of Dijkstra's algorithm for \emph{unweighted} graphs. Similarly to the array of distances $\texttt{dist[]}$, up to two arrays of degeneracies are updated along the algorithm.

We initialise four values:
\begin{itemize}
    \item $l_0$\texttt{[]}, an array of distances from \texttt{source} to each vertex in the \emph{unweighted} contracted syndrome graph. Initially, $l_0\texttt{[source]} = 0$ and $l_0\texttt{[}v\texttt{]} = \infty$ for all the other vertices $v$. As the algorithm progresses, the distance from \texttt{source} to each other vertex is updated.
    \item $Q$, a queue of vertices to be explored. Initially $Q$ contains only \texttt{source}. At the end of the algorithm, $Q$ is empty.
    \item $\Omega_{0}$\texttt{[]}, an array of first-order degeneracies between \texttt{source} and each vertex in the contracted syndrome graph. Initially, $\Omega_{0}\texttt{[source]} = 1$. As the algorithm progresses, $\Omega_{0}$\texttt{[]} between \texttt{source} and each other vertex is updated.
    \item $\Omega_{1}$\texttt{[]}, an array of second-order degeneracies between \texttt{source} and each vertex in the contracted syndrome graph. Initially, $\Omega_{1}\texttt{[}v\texttt{]} = 0$ for all vertices $v$. As the algorithm progresses, $\Omega_{1}$\texttt{[]} between \texttt{source} and each other vertex is updated.
\end{itemize}
Notice that $Q$ here is a normal queue, with only two operation:  \texttt{add\_end()} and \texttt{extract\_first\_element()}. The operation \texttt{add\_end($v$)} adds $v$ at the end of the queue, and the operation \texttt{extract\_first\_element()} extracts the first element in the queue.

The Dijkstra's algorithm for unweighted graphs is slightly different. This is because we need to update the distance of a given vertex only once. In summary, the algorithm first initialises the array of distances, the two arrays of degeneracies and the queue. Then, at each step, the first vertex of $Q$ is extracted and set as the \texttt{current} vertex. We consider all its children. There are up to four cases we need to analyse:
\begin{enumerate}
    \item $l_0\texttt{[child]} = \infty$: $\texttt{child}$ has not been visited, so its distance is updated to $l_0\texttt{[current]} + 1$. We update $\Omega_{0}\texttt{[child]}$ to $\Omega_{0}\texttt{[current]}\cdot \omega(\texttt{current},\texttt{child})$, where $\omega(e)$ is the time overlap for edge $e$. The vertex \texttt{child} is then added to the end of $Q$.
    \item $l_0\texttt{[child]} = l_0\texttt{[current]} + 1$: \texttt{child} has been visited before through a different shortest path. We update $\Omega_{0}\texttt{[child]}$ by adding $\Omega_{0}\texttt{[current]}\cdot \omega(\texttt{current},\texttt{child})$ to its old value.
    \item $l_0\texttt{[child]} = l_0\texttt{[current]}$: \texttt{child} and \texttt{current} are equidistant from \texttt{source}, and the edge (\texttt{child}, \texttt{current}) belongs to a second-shortest path. We update $\Omega_{1}\texttt{[child]}$ by adding $\Omega_{0}\texttt{[current]}\cdot \omega(\texttt{current},\texttt{child})$ to its old value.
    \item $l_0\texttt{[child]} = l_0\texttt{[current]} - 1$: \texttt{current} may be visited through a second-shortest path. We update $\Omega_{1}\texttt{[current]}$ (and not $\Omega_{1}\texttt{[child]}$) by adding $\Omega_{1}\texttt{[child]}\cdot \omega(\texttt{current},\texttt{child})$ to its old value.
\end{enumerate}
We repeat the above procedure until $Q$ is empty.

The update of $\Omega_{1}\texttt{[]}$ works by noticing that Dijkstra's algorithm explores the vertices in layers. First all vertices with distance $1$ are queued and later explored, followed by all vertices with distance $2$, and so on. A second-shortest path can only happen if it is a combination of a shortest path with an edge between two vertices from the same layer, i.e., with the same distance $l_0\texttt{[]}$. Condition~$3$ ($l_0\texttt{[child]} = l_0\texttt{[current]}$) ensures that we go from a shortest path to a second-shortest path via an edge between vertices in the same layer. We then need to use the first-order degeneracy $\Omega_{0}\texttt{[]}$ to update the second-order degeneracy $\Omega_{1}\texttt{[]}$. This works because at this point of the algorithm all values of $\Omega_{0}\texttt{[]}$ for the given layer were already calculated. On the other hand, condition~$4$ ($l_0\texttt{[child]} = l_0\texttt{[current]} - 1$) ensures that we stay on a second-shortest path if there is one through $\texttt{child}$, as the transition between shortest and second-shortest paths happened in some previous layer. Hence the use of a second-order degeneracy $\Omega_{1}\texttt{[]}$ to also update a second-order degeneracy. This works since the values of $\Omega_{1}\texttt{[]}$ for a given layer are completely calculated once all its vertices are considered (differently from $\Omega_{0}\texttt{[]}$, whose values are calculated when all the vertices from the \emph{previous} layer have been considered). Moreover, if there is no second-shortest path through $\texttt{child}$ to $\texttt{current}$, then $\Omega_{1}\texttt{[child]}=0$ and 
so $\Omega_{1}\texttt{[current]}$ is unchanged.


We provide a pseudo-code below for our adapted Dijkstra's algorithm. If we are not required to compute the second-order degeneracy term $\Omega_{1}\texttt{[]}$, then all the lines regarding it can be ignored (lines $4,18,19,20,21$). 

\begin{algorithm}[H]
    \caption{Dijkstra's algorithm for first and second-order degeneracy}
    \KwInput{Syndrome graph $C$ and \texttt{source} vertex}
    \ForAll{\upshape vertex $v\in C$} {
        $l_0$\texttt{[}$v$\texttt{]} $\gets$ INFINITY\;
        $\Omega_{0}\texttt{[}v\texttt{]} \gets$ UNDEFINED\;
        $\Omega_{1}\texttt{[}v\texttt{]} \gets$ 0\;}
    \texttt{$l_0$[source]} $\gets$ 0\;
    \texttt{$\Omega_{0}$[source]} $\gets 1$\;
    create queue $Q$\;
    $Q$.\texttt{add\_end(source)}\;
    \While{$Q \neq \emptyset$} {
        $u \gets Q$.\texttt{extract\_first\_element()}\;
        \For{\upshape each neighbor $v$ of $u$} {
            \uIf{\upshape $l_0$\texttt{[}$v$\texttt{]} = INFINITY} {
                $Q$.\texttt{add\_end($v$)}\;
                $l_0$\texttt{[}$v$\texttt{]} = $l_0$\texttt{[}$u$\texttt{]} + 1\;
                $\Omega_{0}\texttt{[}v\texttt{]} = \Omega_{0}\texttt{[}u\texttt{]}\cdot\omega(u,v)$\; }
            \uElseIf{\upshape $l_0$\texttt{[}$v$\texttt{]} = $l_0$\texttt{[}$u$\texttt{]} + 1} {
                $\Omega_{0}\texttt{[}v\texttt{]} \gets \Omega_{0}\texttt{[}v\texttt{]} + \Omega_{0}\texttt{[}u\texttt{]}\cdot\omega(u,v)$\;
            }
            \uElseIf{\upshape $l_0$\texttt{[}$v$\texttt{]} = $l_0$\texttt{[}$u$\texttt{]}} {
                $\Omega_{1}\texttt{[}v\texttt{]} \gets \Omega_{1}\texttt{[}v\texttt{]} + \Omega_{0}\texttt{[}u\texttt{]}\cdot\omega(u,v)$\;
            }
            \uElseIf{\upshape $l_0$\texttt{[}$v$\texttt{]} = $l_0$\texttt{[}$u$\texttt{]} $-$ 1} {
                $\Omega_{1}\texttt{[}u\texttt{]} \gets \Omega_{1}\texttt{[}u\texttt{]} + \Omega_{1}\texttt{[}v\texttt{]}\cdot\omega(u,v)$\;
            }
        }
    }
    \Return \texttt{$l_0$[]}, \texttt{$\Omega_{0}$[]}, \texttt{$\Omega_{1}$[]}\;
\end{algorithm}
\vfill
\end{document}